\documentclass[12pt]{article}
\usepackage{epsfig}
\begin{document}
\title{Momentum dependence of pion-induced $\phi$ meson production on nuclei near threshold}
\author{E. Ya. Paryev$^{1,2}$\\
{\it $^1$Institute for Nuclear Research, Russian Academy of Sciences,}\\
{\it Moscow 117312, Russia}\\
{\it $^2$Institute for Theoretical and Experimental Physics,}\\
{\it Moscow 117218, Russia}}

\renewcommand{\today}{}
\maketitle

\begin{abstract}
We study the near-threshold pion-induced production of $\phi$ mesons off nuclei
in the kinematical conditions of the HADES experiment, recently performed at GSI.
The calculations have been performed within a collision model based on the nuclear spectral function.
The model accounts for both the primary $\pi^-$ meson--proton
${\pi^-}p \to {\phi}n$ and the secondary pion--nucleon ${\pi}N \to {\phi}N$ $\phi$ production processes
as well as the effects of the nuclear $\phi$ and nucleon mean-field potentials. We find that the
primary reaction channel ${\pi^-}p \to {\phi}n$ dominates in the $\phi$ production off $^{12}$C and $^{184}$W
target nuclei in the HADES acceptance window at incident pion momentum of 1.7 GeV/c.
We calculate the momentum dependence of the absolute and relative (transparency ratio)
$\phi$ meson yields from the above direct channel. The calculations have been performed for
this initial pion momentum allowing for different
options for the ${\phi}N$ absorption cross section $\sigma_{{\phi}N}$
and different scenarios for the in-medium mass shifts of the $\phi$ meson
and secondary neutron, produced together with $\phi$ in this channel.
We demonstrate that the transparency ratio for the $\phi$
mesons has, contrary to the absolute cross sections, an insignificant sensitivity to the $\phi$ meson
and secondary neutron in-medium mass shifts at $\phi$ momenta studied in the HADES experiment.
On the other hand, we show that there are measurable changes in the transparency ratio
due to the ${\phi}N$ absorption cross section, which means that such relative observable can be useful to help
determine this cross section from the data taken in the HADES experiment.
\end{abstract}

\newpage

\section*{1 Introduction}

\hspace{0.5cm} The study of the effective masses and widths of light vector mesons $\rho$, $\omega$ and $\phi$
in nuclear matter through their production and decays, both hadronic and electromagnetic, in the collisions of
photon-, proton- and heavy-ion beams with nuclear targets has received considerable attention in recent years
(see, for example, Refs.~[1--5]). Among these, there is a special interest in the $\phi$ meson [6].
In particular, the $\phi$ meson (in contrast to the $\rho$ and $\omega$ mesons) is a very narrow resonance in
vacuum and does not overlap with other light resonances in the mass spectrum. This could allow for the
measurement of any modifications of its properties in nuclei.
From various approaches,
based on the QCD sum rules [7--10], on the hadronic models [11--14] and on the Quark-Meson Coupling
model [15], the downward mass shift of the $\phi$ meson in nuclear medium at threshold is expected to be
small, about 1--3\% of its free mass at normal nuclear matter density $\rho_0$, but its in-medium total
width is substantially increased, by a factor of ten at density $\rho_0$ compared to the vacuum value
of 4.3 MeV. The 3.4\% $\phi$ mass
reduction and a width increase by a factor of 3.6 at this density have been reported by the
KEK-E325 experiment [16], in which a comparative study of $\phi$ meson production on C and Cu
target nuclei with proton beams of 12 GeV has been performed.

 $\phi$ production in nuclei has also been studied experimentally by the LEPS [17], CLAS [18] and
ANKE [19, 20] Collaborations at the SPring-8, JLab and COSY facilities. Large in-medium
${\phi}N$ absorption cross sections $\sigma_{{\phi}N}$ of about 35 mb and of 16--70 mb were extracted
by Glauber-type analysis from the $\phi$ photoproduction data collected by the LEPS and CLAS
Collaborations [17, 18] for average $\phi$ momenta $\approx$ 1.8 and 2 GeV/c, respectively. This cross
section in the range of 14--21 mb at $\phi$ momenta of 0.6--1.6 GeV/c has been inferred by the ANKE
Collaboration [19, 20] from the comparison of the measured transparency ratio normalized to carbon with
the respective model calculations. These are to be compared to the free value of $\approx$ 10 mb.
$\phi$ creation in high-energy heavy-ion collisions has been investigated by the STAR and ALICE
Collaborations at RHIC and LHC [21, 22]. The HADES Collaboration recently observed   deep subthreshold
$\phi$ production in 1.23 A GeV Au+Au collisions [23]. A surprisingly high $\phi/K^-$ multiplicity ratio
with a value of 0.52$\pm$0.16 was measured. This indicates, on the one hand, possible changes in the
$\phi$ meson properties in the nuclear medium (in particular, a possibly attractive mass shift) and,
on the other hand, that the $\phi$ mesons are a possible source of $K^-$ mesons at SIS energies.

 $\phi$ production in nuclei has also been intensively studied theoretically in
proton--nucleus [19, 20, 24--28] and photon--nucleus [29--33] reactions, with the aim of getting information
on possible modification of the $\phi$ spectral function in cold nuclear matter. Unique information on the
in-medium properties of light vector mesons, complementary to that from nucleus--nucleus, proton--nucleus and
photon--nucleus collisions, can be deduced from pion--nucleus reactions [34, 35]. In an attempt to obtain valuable
information on an effective ${\phi}N$ absorption cross section $\sigma_{{\phi}N}$, the near-threshold
$\pi^-$ meson-induced $\phi$ production from $^{12}$C and $^{184}$W target nuclei at incident pion momentum of
1.7 GeV/c has been investigated by the HADES Collaboration [36] at the GSI pion beam facility [37]. For the first
time, the momentum dependence of the absolute and relative (transparency ratio) $\phi$ meson yields was measured
for these targets over the $\phi$ momentum range of 0.6--1.1 GeV/c at laboratory polar angular domain of
10$^{\circ}$--45$^{\circ}$. In this respect, the main purpose of the present work is to get estimates of these
yields in the kinematical conditions of the HADES experiment [36] in the framework of the collision model, based
on the nuclear spectral function, for incoherent primary $\pi^-$ meson--proton ${\pi^-}p \to {\phi}n$ and
secondary pion--nucleon ${\pi}N \to {\phi}N$ $\phi$ production processes in different scenarios for the
${\phi}N$ absorption cross section $\sigma_{{\phi}N}$. In view of the expected data from this experiment, the
estimates can be used as an important tool for determining the in-medium $\phi$--nucleon absorption cross section.

\section*{2 Framework}

\section*{2.1 Direct  $\phi$ meson production mechanism}

\hspace{0.5cm} The direct production of $\phi$ mesons in the kinematical conditions of the HADES experiment
in $\pi^-A$ ($A=^{12}$C and $^{184}$W) collisions at incident momentum of 1.7 GeV/c
can occur in the following $\pi^-p$ elementary process with zero pions in the final state
\footnote{$^)$The free threshold momentum for this process is 1.559 GeV/c.
We can ignore the processes $\pi^-N \to {\phi}N{\pi}$ with one pion in the final state
at the incident momentum of interest, because this momentum is less than their production
threshold momenta in free $\pi^-N$ interactions. Thus, for example, the threshold momentum of the
channel $\pi^-p \to {\phi}n{\pi^0}$ is 1.852 GeV/c. This momentum is larger than the incident pion
momentum of 1.7 GeV/c. Consequently, the processes $\pi^-N \to {\phi}N{\pi}$ are energetically repressed.}$^)$,
\begin{equation}
\pi^-+p \to \phi+n.
\end{equation}
Following Refs. [38, 39], we proxy the in-medium local effective masses $m^*_{h}(r)$
of the final neutron and $\phi$ meson,
participating in the production process (1), with their average in-medium masses $<m^*_{h}>$
defined as:
\begin{equation}
<m^*_{h}>=m_{h}+U_h\frac{<{\rho_N}>}{{\rho_0}}.
\end{equation}
Here, $m_{h}$ is the rest mass of a hadron in free space, $U_h$ is the hadron effective scalar
nuclear potential (or its in-medium mass shift) at normal nuclear matter
density ${\rho_0}$, and $<{\rho_N}>$ is the average nucleon density.
For target nuclei $^{12}$C and $^{184}$W,  the ratio $<{\rho_N}>/{\rho_0}$, as shown in our
calculations, is equal to 0.55 and 0.76, respectively. We will use these values in our subsequent study.
The total energy $E^\prime_{h}$ of the hadron in nuclear matter is
expressed via its average effective mass $<m^*_{h}>$ and its in-medium momentum
${\bf p}^{\prime}_{h}$ by the equation:
\begin{equation}
E^\prime_{h}=\sqrt{({\bf p}^{\prime}_{h})^2+(<m^*_{h}>)^2}.
\end{equation}
The momentum ${\bf p}^{\prime}_{h}$ is related to the vacuum hadron momentum ${\bf p}_{h}$
as follows:
\begin{equation}
E^\prime_{h}=\sqrt{({\bf p}^{\prime}_{h})^2+(<m^*_{h}>)^2}=
\sqrt{{\bf p}^2_{h}+m^2_{h}}=E_h,
\end{equation}
where $E_h$ is the hadron total energy in vacuum.

In view of the substantial uncertainties of the $\phi$ meson self-energy at finite momenta [14],
in the rest of this work it is natural to use
the value corresponding to a 2\% reduction of its mass at saturation density $\rho_0$, namely
$U_{\phi}=-20$ MeV for the $\phi$ mass shift $U_{\phi}$ at these momenta.
The effective scalar nucleon potential $U_N$, used in
Eq.~(2), can be determined from the relation
\begin{equation}
U_N=\frac{\sqrt{m_N^2+{p^{\prime}_N}^2}}{m_N}V_{NA}^{\rm SEP},
\end{equation}
where $V_{NA}^{\rm SEP}$ is the Schr${\ddot{\rm o}}$dinger equivalent potential
for nucleons at normal nuclear matter density.
Accounting for the fact that in the kinematics of the HADES experiment
the momenta $p^{\prime}_N$ of the final neutron participating in reaction (1)
are, as shown in our calculations, around a momentum of 0.8 GeV/c
and using $V_{NA}^{\rm SEP} \approx 20$ MeV at this momentum, corresponding to a kinetic energy
\footnote{$^)$For a free particle dispersion relation.}$^)$
of 0.3 GeV [40], we obtain that $U_N \approx 25$ MeV.
We will adopt this potential in the rest of this paper. In order to see the sensitivity of the
$\phi$ production cross sections from the one-step process (1) to the $\phi$ and neutron mass shifts
$U_{\phi}$ and $U_N$, we will also ignore them in our calculations.

  Because the $\phi$--nucleon elastic cross section is expected to be small for the $\phi$ momenta of
0.6--1.1 GeV/c studied in the HADES experiment [34, 41], we will neglect quasielastic ${\phi}N$
rescatterings in the present study. Then, taking into consideration
the distortion of the incident pion in nuclear matter and describing the $\phi$ meson final-state
absorption by the effective in-medium cross section $\sigma_{{\phi}N}$
as well as using the results presented in Refs.~[27, 39, 42], we represent the inclusive differential
cross section for the production of ${\phi}$ mesons with vacuum momentum ${\bf p}_{\phi}$
off nuclei in the direct process (1) as follows:
\begin{equation}
\frac{d\sigma_{{\pi^-}A\to {\phi}X}^{({\rm prim})}
({\bf p}_{\pi^-},{\bf p}_{\phi})}
{d{\bf p}_{\phi}}=I_{V}[A,\theta_{\phi}]
\left(\frac{Z}{A}\right)\left<\frac{d\sigma_{{\pi^-}p\to {\phi}n}({\bf p}_{\pi^-},
{\bf p}^{\prime}_{\phi})}{d{\bf p}^{\prime}_{\phi}}\right>_A
\frac{d{\bf p}^{\prime}_{{\phi}}}{d{\bf p}_{{\phi}}},
\end{equation}
where
\begin{equation}
I_{V}[A,\theta_{\phi}]=A\int\limits_{0}^{R}r_{\bot}dr_{\bot}
\int\limits_{-\sqrt{R^2-r_{\bot}^2}}^{\sqrt{R^2-r_{\bot}^2}}dz
\rho(\sqrt{r_{\bot}^2+z^2})
\exp{\left[-\sigma_{{\pi^-}N}^{\rm tot}A\int\limits_{-\sqrt{R^2-r_{\bot}^2}}^{z}
\rho(\sqrt{r_{\bot}^2+x^2})dx\right]}
\end{equation}
$$
\times
\int\limits_{0}^{2\pi}d{\varphi}\exp{\left[-\sigma_{{\phi}N}A\int\limits_{0}^{l(\theta_{\phi},\varphi)}
\rho(\sqrt{x^2+2a(\theta_{\phi},\varphi)x+b+R^2})dx\right]},
$$
\begin{equation}
a(\theta_{\phi},\varphi)=z\cos{\theta_{\phi}}+r_{\bot}\sin{\theta_{\phi}}\cos{\varphi},\\\
b=r_{\bot}^2+z^2-R^2,
\end{equation}
$$
l(\theta_{\phi},\varphi)=\sqrt{a^2(\theta_{\phi},\varphi)-b}-
a(\theta_{\phi},\varphi),
$$
\begin{equation}
\left<\frac{d\sigma_{{\pi^-}p\to {\phi}n}({\bf p}_{\pi^-},{\bf p}_{\phi}^{'})}
{d{\bf p}_{\phi}^{'}}\right>_A=
\int\int
P_A({\bf p}_t,E)d{\bf p}_tdE
\end{equation}
$$
\times
\left\{\frac{d\sigma_{{\pi^-}p\to {\phi}n}[\sqrt{s},<m^*_{{\phi}}>,
<m^*_{N}>,{\bf p}_{\phi}^{'}]}
{d{\bf p}_{\phi}^{'}}\right\}
$$
and
\begin{equation}
  s=(E_{\pi^-}+E_t)^2-({\bf p}_{\pi^-}+{\bf p}_t)^2,
\end{equation}
\begin{equation}
   E_t=M_A-\sqrt{(-{\bf p}_t)^2+(M_{A}-m_{N}+E)^{2}}.
\end{equation}
Here,
$d\sigma_{{\pi^-}p\to {\phi}n}[\sqrt{s},<m^*_{{\phi}}>,<m^*_{N}>,{\bf p}_{\phi}^{'}] /d{\bf p}_{\phi}^{'}$
is the off-shell inclusive differential cross section for the production of ${\phi}$ mesons and neutrons
with reduced masses $<m^*_{{\phi}}>$ and $<m^*_{N}>$, respectively, and $\phi$ mesons with in-medium momentum ${\bf p}_{{\phi}}^{'}$ in reaction (1) at the ${\pi^-}p$ center-of-mass energy $\sqrt{s}$;
$E_{\pi^-}$ and ${\bf p}_{\pi^-}$ are the total energy and momentum of the incident pion
($E_{\pi^-}=\sqrt{m^2_{\pi}+{\bf p}^2_{\pi^-}}$, $m_{\pi}$ is the free space pion mass);
$\rho({\bf r})$ and $P_A({\bf p}_t,E)$ are the local nucleon density and the
spectral function of the target nucleus A normalized to unity;
${\bf p}_t$ and $E$ are the internal momentum and removal energy of the struck target proton
involved in the collision process (1); $\sigma_{{\pi^-}N}^{\rm tot}$ is the total cross section of the
free ${\pi^-}N$ interaction; $Z$ and $A$ are the numbers of protons and nucleons in
the target nucleus, and $M_{A}$  and $R$ are its mass and radius; and $\theta_{\phi}$ is the polar angle of
vacuum momentum ${\bf p}_{{\phi}}$ in the laboratory system with z-axis directed along the momentum
${\bf p}_{{\pi^-}}$ of the initial pion. Since the momenta of the outgoing neutrons in reaction (1)
in the kinematics of the HADES experiment are substantially greater than the typical average Fermi
momentum of $\sim$ 250 MeV/c of the target nucleus (see below),
we neglect the correction of Eq.~(9) for
the Pauli blocking, leading to suppression of the phase space available for them
\footnote{$^)$We have checked that it has no essential influence on our results
at our incident pion momentum of interest, 1.7 GeV/c.}$^)$
.

        For the nuclear density $\rho({\bf r})$ in the cases of the $^{12}$C and $^{184}$W
target nuclei considered, we have adopted, respectively, the harmonic oscillator
and the Woods-Saxon distributions:
\begin{equation}
\rho({\bf r})={\rho}_{N}({\bf r})/A=\frac{(b/\pi)^{3/2}}{A/4}\left\{1+
\left[\frac{A-4}{6}\right]br^{2}\right\}\exp{(-br^2)},
\end{equation}
\begin{equation}
 \rho({\bf r})=\rho_{0}\left[1+
\exp{\left(\frac{r-R_{1/2}}{a}\right)}\right]^{-1},
\end{equation}
where $b=0.355~{\rm fm}^{-2}$, $R_{1/2}=6.661~{\rm fm}$ and $a=0.55~{\rm fm}$.
For the $\phi$ production calculations in the case of the $^{12}$C target nucleus
the nuclear spectral function $P_A({\bf p}_t,E)$ was taken from Ref.~[43].
For the $^{184}$W target nucleus its single-particle part was assumed to be the same
as that for $^{208}$Pb [44]. This latter was taken from Ref.~[45]. The correlated part of the
spectral function for $^{184}$W was taken from Ref.~[43].

   Following Refs. [39, 42], we suppose that the off-shell inclusive differential cross section\\
$d\sigma_{{\pi^-}p\to {\phi}n}[\sqrt{s},<m^*_{{\phi}}>,<m^*_{N}>,{\bf p}_{\phi}^{'}] /d{\bf p}_{\phi}^{'}$
for $\phi$ production in reaction (1) is equivalent to the respective on-shell cross section calculated for
the off-shell kinematics of this reaction as well as for the final $\phi$ and neutron in-medium masses
$<m^*_{{\phi}}>$ and $<m^*_{N}>$.
    Accounting for Eq.~(16) from Ref. [39], we get the following expression for the
elementary in-medium differential cross section
$d\sigma_{{\pi^-}p\to {\phi}n}[\sqrt{s},<m^*_{{\phi}}>,<m^*_{N}>,{\bf p}_{\phi}^{'}] /d{\bf p}_{\phi}^{'}$:
\begin{equation}
\frac{d\sigma_{{\pi^{-}}p \to {\phi}n}[\sqrt{s},<m^*_{\phi}>,<m^*_{N}>,
{\bf p}^{\prime}_{\phi}]}{d{\bf p}^{\prime}_{\phi}}=
\frac{\pi}{I_2[s,<m^*_{\phi}>,<m^*_{N}>]E^{\prime}_{\phi}}
\end{equation}
$$
\times
\frac{d\sigma_{{\pi^{-}}p \to {\phi}n}(\sqrt{s},<m^*_{\phi}>,<m^*_{N}>,\theta^*_{\phi})}
{d{\bf \Omega}^*_{\phi}}
$$
$$
\times
\frac{1}{(\omega+E_t)}\delta\left[\omega+E_t-\sqrt{(<m^*_{N}>)^2+({\bf Q}+{\bf p}_t)^2}\right],
$$
where
\begin{equation}
I_2[s,<m^*_{\phi}>,<m^*_{N}>]=\frac{\pi}{2}
\frac{\lambda[s,(<m^*_{\phi}>)^{2},(<m^*_{{N}}>)^{2}]}{s},
\end{equation}
\begin{equation}
\lambda(x,y,z)=\sqrt{{\left[x-({\sqrt{y}}+{\sqrt{z}})^2\right]}{\left[x-
({\sqrt{y}}-{\sqrt{z}})^2\right]}},
\end{equation}
\begin{equation}
\omega=E_{\pi^-}-E^{\prime}_{\phi}, \,\,\,\,{\bf Q}={\bf p}_{\pi^-}-{\bf p}^{\prime}_{\phi}.
\end{equation}
Here,
$d\sigma_{{\pi^{-}}p \to {\phi}n}(\sqrt{s},<m^*_{\phi}>,<m^*_{N}>,\theta^*_{\phi})/d{\bf \Omega}^*_{\phi}$
is the off-shell differential cross section for the production of $\phi$ mesons with mass $<m^*_{\phi}>$
in reaction (1) under the polar angle $\theta^*_{\phi}$ in the ${\pi^-}p$ c.m.s., which is assumed to be
isotropic [27, 34] in our calculations of $\phi$ meson creation in ${\pi^-}A$ collisions from this reaction:
\begin{equation}
\frac{d\sigma_{{\pi^{-}}p \to {\phi}n}(\sqrt{s},<m^*_{\phi}>,<m^*_{N}>,\theta^*_{\phi})}
{d{\bf \Omega}^*_{\phi}}=\frac{\sigma_{{\pi^{-}}p \to {\phi}n}(\sqrt{s},\sqrt{s^*_{\rm th}})}{4\pi}.
\end{equation}
Here, $\sigma_{{\pi^{-}}p \to {\phi}n}(\sqrt{s},\sqrt{s^*_{\rm th}})$ is the ``in-medium" total cross section
of reaction (1) having the threshold energy $\sqrt{s^*_{\rm th}}=<m^*_{\phi}>+<m^*_{N}>$.
In line with the above, it is equivalent to the vacuum cross section
$\sigma_{{\pi^{-}}p \to {\phi}n}(\sqrt{s},\sqrt{s_{\rm th}})$, in which the free threshold energy
$\sqrt{s_{\rm th}}=m_{\phi}+m_{n}=1.959$ GeV is replaced by the in-medium threshold
energy $\sqrt{s^*_{\rm th}}$. For the free total cross section
$\sigma_{{\pi^{-}}p \to {\phi}n}(\sqrt{s},\sqrt{s_{\rm th}})$ we have used the following parametrization
suggested in Ref. [46]:
\begin{equation}
\sigma_{{\pi}^-p \to {\phi}n}(\sqrt{s},\sqrt{s_{\rm th}})=\left\{
\begin{array}{ll}
	0.47\left(\sqrt{s}-\sqrt{s_{\rm th}}\right)~[{\rm mb}]
	&\mbox{for $\sqrt{s_{\rm th}}<\sqrt{s}< 2.05~{\rm GeV}$}, \\
	&\\
                   23.7/s^{4.4}~[{\rm mb}]
	&\mbox{for $\sqrt{s}\ge 2.05~{\rm GeV}$}.
\end{array}
\right.	
\end{equation}
Using Eq.~(19), one can easily obtain that the cross section $\sigma_{{\pi}^-p \to {\phi}n}$
amounts to 31 $\mu$b for the initial pion momentum of 1.7 GeV/c, corresponding to center-of-mass energy
$\sqrt{s}=2.025$ GeV or to excess energy $\sqrt{s}-\sqrt{s_{\rm th}}=66$ MeV.
It should be noted that for the cross section $\sigma_{{\pi}^{-}p \to {\phi}n}$ we have also adopted another parametrization:
\begin{equation}
\sigma_{{\pi}^-p \to {\phi}n}(\sqrt{s},\sqrt{s_{\rm th}})=\left\{
\begin{array}{ll}
	0.101\left(\sqrt{s}-\sqrt{s_{\rm th}}\right)^{0.466}~[{\rm mb}]
	&\mbox{for $\sqrt{s_{\rm th}}<\sqrt{s}< 2.08~{\rm GeV}$}, \\
	&\\
                   23.7/s^{4.4}~[{\rm mb}]
	&\mbox{for $\sqrt{s}\ge 2.08~{\rm GeV}$},
\end{array}
\right.	
\end{equation}
which combines the relevant parametrization from Ref. [47] in the ``low"-energy region
with that given by Eq. (19) in the ``high"-energy region. The parametrization (20) is close to the results from the boson-exchange model [48] near the threshold, where the data are not available,
and is considerably larger here (at $\sqrt{s}-\sqrt{s_{\rm th}} \le 10$ MeV) than the one given above by Eq. (19).
However, for pion beam momentum of 1.7 GeV/c, Eq.~(20) gives for the cross section
$\sigma_{{\pi}^{-}p \to {\phi}n}$ the value of 28.5 $\mu$b, which is close to that obtained above at this momentum
by means of Eq. (19). As shown in our calculations, the differences between momentum differential cross sections
for $\phi$ meson production from the primary ${\pi}^{-}p \to {\phi}n$ channel on the considered target nuclei,
obtained by using different parametrizations (19) and (20) in the kinematical conditions of interest,
are insignificant. They are within a few percent.
Therefore, this gives us confidence that Eq.~(19) is reliable enough
to describe $\phi$ production on nuclei through the ${\pi^-}p \to {\phi}n$ reaction
\footnote{$^)$And via the ${\pi}N \to {\phi}N$ processes, see below.}$^)$
.
We will employ this expression throughout our calculations.

We define now the cross sections $\sigma_{{\pi^-}N}^{\rm tot}$
and $\sigma_{{\phi}N}$ in Eq.~(7). We adopt $\sigma_{{\pi^-}N}^{\rm tot}=35$ mb
for the incident pion momentum of 1.7 GeV/c [49]. For the cross section
$\sigma_{{\phi}N}$  we also adopt the values  $\sigma_{{\phi}N}=10,20,30$, and 40 mb motivated by the
results from the $\phi$ photoproduction experiments at SPring-8/Osaka [17], JLab [18, 50], from the
proton--nucleus ANKE-at-COSY experiment [19, 20], from the analysis [32] of the data [17]
within the GiBUU transport model, from the Vector Dominance Model analysis [51] of the forward
${\gamma}p \to {\phi}p$ differential cross section, from the analysis in this work of coherent and
incoherent $\phi$ meson photoproduction on nuclei in terms of single and coupled-channel photoproduction,
as well as from the estimates [52] for $\sigma_{{\phi}N}$ in vacuum and in matter.
In Eqs.~(7) and (8) we assume that the
direction of the rather high three-momentum of the $\phi$ meson
remains unchanged as it propagates from its production point inside the nucleus in
weak nuclear mean-field, considered in this paper, to the vacuum outside the nucleus.
In this case, the quantities
$\left<d\sigma_{{\pi^-}p\to {\phi}n}({\bf p}_{\pi^-},
{\bf p}^{\prime}_{\phi})/d{\bf p}^{\prime}_{\phi}\right>_A$
and
$d{\bf p}^{\prime}_{\phi}/d{\bf p}_{\phi}$,
used in Eq. (6), can be taken into account in our calculations as
$\left<d\sigma_{{\pi^-}p\to {\phi}n}(p_{\pi^-},
p^{\prime}_{\phi},\theta_{\phi})/p^{\prime2}_{\phi}dp^{\prime}_{\phi}d{\bf \Omega}_{\phi}\right>_A$
and
$p^{\prime}_{\phi}/{p}_{\phi}$, where
${\bf \Omega}_{\phi}(\theta_{\phi},\varphi_{\phi})={\bf p}_{\phi}/p_{\phi}$. Here, $\varphi_{\phi}$ is the
azimuthal angle of the $\phi$ momentum ${\bf p}_{\phi}$ in the lab system. With these, and
integrating the full inclusive differential cross section (6) over the HADES acceptance window
${\Delta}{\bf \Omega}_{\phi}$=$10^{\circ} \le \theta_{\phi} \le 45^{\circ}$ and
$0 \le \varphi_{\phi} \le 2{\pi}$, we can represent the differential cross section for $\phi$ meson
production in ${\pi^-}A$ collisions from the one-step process (1), corresponding to the kinematical
conditions of the HADES experiment, in the following form:
\begin{equation}
\frac{d\sigma_{{\pi^-}A\to {\phi}X}^{({\rm prim})}
(p_{\pi^-},p_{\phi})}{dp_{\phi}}=\int\limits_{{\Delta}{\bf \Omega}_{\phi}}^{}d{\bf \Omega}_{\phi}
\frac{d\sigma_{{\pi^-}A\to {\phi}X}^{({\rm prim})}
({\bf p}_{\pi^-},{\bf p}_{\phi})}{d{\bf p}_{\phi}}p_{\phi}^2
\end{equation}
$$
=2{\pi}\left(\frac{Z}{A}\right)\left(\frac{p_{\phi}}{p^{\prime}_{\phi}}\right)
\int\limits_{\cos45^{\circ}}^{\cos10^{\circ}}d\cos{{\theta_{\phi}}}I_{V}[A,\theta_{\phi}]
\left<\frac{d\sigma_{{\pi^-}p\to {\phi}n}(p_{\pi^-},
p^{\prime}_{\phi},\theta_{\phi})}{dp^{\prime}_{\phi}d{\bf \Omega}_{\phi}}\right>_A.
$$

  If the primary pion-induced reaction channel (1) dominates in the $\phi$ production in ${\pi^-}A$
interactions near threshold, then the absorption cross section $\sigma_{{\phi}N}$
(and in principle its momentum dependence) can be extracted, in particular, from a comparison of the
momentum distributions
calculated on the basis of Eq. (21) and measured momentum distributions of $\phi$ mesons produced off
$^{12}$C and $^{184}$W target nuclei by pions with momentum of 1.7 GeV/c.
An alternative way to estimate the inelastic in-medium $\phi$--nucleon cross section $\sigma_{{\phi}N}$
would be through a direct fit of the measured relative transparency ratio
by the following relative observable -- the
transparency ratio $T_A$ defined as the ratio between the inclusive differential $\phi$ production
cross section (21) on a heavy nucleus ($^{184}$W) and on a light one ($^{12}$C), viz.:
\begin{equation}
T_A(p_{\pi^-},p_{\phi})=\frac{12}{A}\frac{d\sigma_{{\pi^-}A\to {\phi}X}^{({\rm prim})}
(p_{\pi^-},p_{\phi})/dp_{\phi}}
{d\sigma_{{\pi^-}{\rm C}\to {\phi}X}^{({\rm prim})}(p_{\pi^-},p_{\phi})/dp_{\phi}}.
\end{equation}
Such a relative observable is more favorable than those like Eq.~(21) based on the
absolute cross sections for the aim of getting the information on meson in-medium absorption cross section
both from the experimental and theoretical sides. On the one hand, it
allows for a reduction of systematic errors due to the cancellation of the efficiency corrections
and, on the other hand, the theoretical uncertainties associated with the meson production mechanism
(in particular, with meson production cross section off the neutron) substantially cancel out.

\section*{2.2 Two-step ${\phi}$ production mechanisms}

\hspace{0.5cm} The kinematical arguments lead to the suggestion that the following two-step production
processes with a pion in an intermediate states may contribute to  ${\phi}$ production
in ${\pi^-}A$ interactions at the initial momentum of interest. An incident $\pi^-$ meson scatters
quasielastically on target nucleon (proton or neutron)
\begin{equation}
\pi^-+p \to \pi^- +p,
\end{equation}
\begin{equation}
\pi^-+n \to \pi^- +n,
\end{equation}
in such a way that the phase space for the final low-momentum nucleon is not Pauli blocked and the final
$\pi^-$ meson (which is assumed to be on-shell)
turns out to be able to create a $\phi$ meson via the elementary process (1)
\footnote{$^)$Our estimate shows that at pion beam momentum of 1.7 GeV/c and for free target nucleons at rest,
involved in processes (1), (23) and (24) as well as for typical average Fermi momentum
${\bar p}_F \sim$ 250 MeV/c, these two conditions are fulfilled if the incident pion scatters in the
polar angular domain of $10^{\circ} \le \theta_{\pi^-} \le 20^{\circ}$ in the lab system.}$^)$
.
An incident pion produces,
in the first inelastic collision with an intranuclear nucleon, two $\pi$ mesons and a nucleon via the
elementary reactions:
\begin{equation}
\pi^-+p \to \pi^- +\pi^0+p,
\end{equation}
\begin{equation}
\pi^-+p \to \pi^+ +\pi^-+n;
\end{equation}
\begin{equation}
\pi^-+n \to \pi^- +\pi^0+n,
\end{equation}
\begin{equation}
\pi^-+n \to \pi^- +\pi^-+p.
\end{equation}
Then, the most energetic secondary pion ($\pi^-$, $\pi^0$, $\pi^+$), which is also assumed to be on-shell,
creates the $\phi$ meson on another nucleon of the target nucleus both via the subprocess (1) and through
the elementary channels
\begin{equation}
\pi^++n \to \phi +p,
\end{equation}
\begin{equation}
\pi^0+p \to \phi +p,
\end{equation}
\begin{equation}
\pi^0+n \to \phi +n,
\end{equation}
provided that these channels are allowed energetically. Thus, for example, at a pion beam momentum
of 1.7 GeV/c the maximum allowable momentum of a secondary pion in the process ${\pi^-}N \to 2{\pi}N$,
proceeding on a free nucleon at rest, is $\approx$ 1.55 GeV/c. This momentum is close to the threshold
momentum of 1.56 GeV/c of the channel ${\pi}N \to {\phi}N$. Consequently, accounting for the Fermi motion of
intranuclear nucleons, this channel may contribute to the $\phi$ production on nuclei. In our calculations
of this production we have neglected the elementary processes ${\pi^-}p \to {\pi^0}n$,
${\pi^-}p \to 2{\pi^0}n$, ${\pi^-}N \to 3{\pi}N$, since their cross sections
at a beam momentum of 1.7 GeV/c are expected to be
essentially less [49] than those of reactions (23), (24) and (25)--(28), which are shown
(for free space) in Figs. 1--6 given below.
For the four-body channel ${\pi^-}N \to 3{\pi}N$, the produced pions are less energetic
than those from three-body reactions ${\pi^-}N \to 2{\pi}N$. Thus, at a beam
momentum of 1.7 GeV/c the maximum allowable momentum of a pion in this channel, taking place on a free
nucleon at rest, is 1.39 GeV/c. This momentum is less than the momentum of 1.55 GeV/c of the latter
reactions.

   In the following calculations we will include the medium modification of the final low-momentum
nucleons, participating in the production processes (23), (24) and (25)--(28) together with high-momentum
pions producing the observed $\phi$ mesons, by using their average in-medium mass $<m_N^{**}>$ defined above
by Eq. (2) with the potential depth at saturation density $U_N=-60$ MeV, corresponding to their average
momentum $p_N^{\prime}$ $\sim {\bar p}_F \sim 250$ MeV/c [40]. Accounting for the medium effects on the
outgoing $\phi$ mesons and nucleons in the secondary processes (1), (29)--(31)
in the same way as that adopted in calculating
the $\phi$ production cross section (6) from the direct channel (1) and using the results given in Refs. [39, 42],
we get the following expression for the $\phi$ differential production cross section for ${\pi^-}A$
reactions from these processes:
\begin{equation}
\frac{d\sigma_{{\pi^-}A\to {\phi}X}^{({\rm sec})}
({\bf p}_{\pi^-},{\bf p}_{\phi})}{d{\bf p}_{\phi}}=I_{V}^{({\rm sec})}[A]
\sum_{\pi'=\pi^-,\pi^0,\pi^+}\int d{\bf p}_{\pi}
\end{equation}
$$
\times
\left[\frac{Z}{A}\left<\frac{d\sigma_{{\pi^-}p\to {\pi'}X}({\bf p}_{\pi^-},
{\bf p}_{\pi})}{d{\bf p}_{\pi}}\right>_A+
\frac{N}{A}\left<\frac{d\sigma_{{\pi^-}n\to {\pi'}X}({\bf p}_{\pi^-},
{\bf p}_{\pi})}{d{\bf p}_{\pi}}\right>_A\right]
$$
$$
\times
\left[\frac{Z}{A}\left<\frac{d\sigma_{{\pi'}p\to {\phi}N}({\bf p}_{\pi},{\bf p}_{\phi}^{\prime})}
{d{\bf p}_{\phi}^{\prime}}\right>_A+
\frac{N}{A}\left<\frac{\sigma_{{\pi'}n\to {\phi}N}({\bf p}_{\pi},{\bf p}_{\phi}^{\prime})}
{d{\bf p}_{\phi}^{\prime}}\right>_A\right]\frac{d{\bf p}^{\prime}_{{\phi}}}{d{\bf p}_{{\phi}}},
$$
where
\begin{equation}
I_{V}^{({\rm sec})}[A]=2{\pi}A^2\int\limits_{0}^{R}r_{\bot}dr_{\bot}
\int\limits_{-\sqrt{R^2-r_{\bot}^2}}^{\sqrt{R^2-r_{\bot}^2}}dz
\rho(\sqrt{r_{\bot}^2+z^2})
\int\limits_{0}^{\sqrt{R^2-r_{\bot}^2}-z}dl
\rho(\sqrt{r_{\bot}^2+(z+l)^2})
\end{equation}
$$
\times
\exp{\left[-\sigma_{{\pi'}N}^{\rm tot}A\int\limits_{z}^{z+l}
\rho(\sqrt{r_{\bot}^2+x^2})dx
-\sigma_{{\phi}N}A\int\limits_{z+l}^{\sqrt{R^2-r_{\bot}^2}}
\rho(\sqrt{r_{\bot}^2+x^2})dx\right]}
$$
$$
\times
\exp{\left[-\sigma_{{\pi^-}N}^{\rm tot}A\int\limits_{-\sqrt{R^2-r_{\bot}^2}}^{z}
\rho(\sqrt{r_{\bot}^2+x^2})dx\right]},
$$
\begin{equation}
\left<\frac{d\sigma_{{\pi^-}N\to {\pi'}X}({\bf p}_{\pi^-},
{\bf p}_{\pi})}
{d{\bf p}_{\pi}}\right>_A=
\int\int
P_A({\bf p}_t,E)d{\bf p}_tdE
\left\{\frac{d\sigma_{{\pi^-}N\to {\pi'}X}(\sqrt{s},<m_N^{**}>,{\bf p}_{\pi})}
{d{\bf p}_{\pi}}\right\}.
\end{equation}
Here, $d\sigma_{{\pi^-}N\to {\pi'}X}(\sqrt{s},<m_N^{**}>,{\bf p}_{\pi})/d{\bf p}_{\pi}$ are the
in-medium inclusive differential cross sections for pion ``production" from
the primary pion-induced reaction channels (23), (24) and (25)--(28) at the ${\pi^-}N$
center-of-mass energy $\sqrt{s}$ defined above by Eq. (10);
$\sigma_{{\pi'}N}^{{\rm tot}}$ is the total cross section of the free ${\pi'}N$
interaction
\footnote{$^)$It is natural to use
$\sigma_{{\pi'}N}^{{\rm tot}}$=$\sigma_{{\pi^-}N}^{{\rm tot}}$=35 mb in the following calculations.}$^)$
;
${\bf p}_{\pi}$ is the three-momentum of an intermediate pion, which has the total energy
$E_{\pi}=\sqrt{m_{\pi}^2+{\bf p}_{\pi}^2}$; N is the number of neutrons in the target nucleus; and
$<d\sigma_{{\pi'}N\to {\phi}N}({\bf p}_{\pi},{\bf p}_{\phi}^{'})/d{\bf p}_{\phi}^{'}>_A$
are the in-medium inclusive removal energy-internal momentum-averaged differential cross sections
for $\phi$ meson production in the secondary reactions (1), (29)--(31). These quantities are defined
above by Eqs. (9)--(11) and (14)--(18), in which one has to make the following substitutions:
$\sigma_{{\pi^-}p \to {\phi}n}$ $\to$ $\sigma_{{\pi^-}p \to {\phi}n}$, $\sigma_{{\pi^+}n \to {\phi}p}$,
$\sigma_{{\pi^0}p \to {\phi}p}$, $\sigma_{{\pi^0}n \to {\phi}n}$;
${\bf p}_{\pi^-}$ $\to$ ${\bf p}_{\pi}$, $E_{\pi^-}$ $\to$ $E_{\pi}$.
Due to isospin symmetry, the following relations exist among the latter four total cross sections
 [27]:
\begin{equation}
\sigma_{{\pi}^{+}n \to {\phi}p}=\sigma_{{\pi}^{-}p \to {\phi}n},
\end{equation}
\begin{equation}
\sigma_{{\pi}^{0}p \to {\phi}p}=\sigma_{{\pi}^{0}n \to {\phi}n}=\frac{1}{2}\sigma_{{\pi}^{-}p \to {\phi}n}.
\end{equation}
For the free total cross section
$\sigma_{{\pi}^{-}p \to {\phi}n}$ we have used the parametrization (19) given above.

 In the expression (33), which defines the quantity $I_{V}^{({\rm sec})}[A]$, the first two eikonal
factors account for the distortion of the intermediate pion and detected $\phi$ meson on their ways out
of the nucleus, whereas the latter describes the attenuation of the incident pion inside the target
nucleus. Due to the weak dependence of this quantity on the relevant intermediate pion and observed
$\phi$ meson production angles in the laboratory rest frame, we suppose,
writing Eq.~(33), that they move in the nucleus in the forward direction.

 In our approach the differential cross sections
$d\sigma_{{\pi^-}N\to {\pi'}X}(\sqrt{s},<m_N^{**}>,{\bf p}_{\pi})/d{\bf p}_{\pi}$ for pion production
in ${\pi^-}N$ collisions (23), (24) and (25)--(28) have been described by the two- and three-body phase
space calculations, corrected for Pauli blocking, leading to the suppression of the phase space available
for the final-state nucleon. In line with Eqs. (14)--(17), (23), (24) and (25)--(28)
as well as (36)--(48) from Ref. [53], these cross sections can be written in the following forms:
\begin{equation}
\frac{d\sigma_{{\pi^-}p\to {\pi^-}X}(\sqrt{s},<m_N^{**}>,{\bf p}_{\pi})}
{d{\bf p}_{\pi}}=
\frac{d\sigma_{{\pi^-}p\to {\pi^-}p}(\sqrt{s},<m_N^{**}>,{\bf p}_{\pi})}
{d{\bf p}_{\pi}}
\end{equation}
$$
+
\left[\sigma_{{\pi^-}p \to {\pi^-}{\pi^0}p}(\sqrt{s})+
\sigma_{{\pi^-}p \to {\pi^+}{\pi^-}n}(\sqrt{s})\right]f_3(\sqrt{s},<m_N^{**}>,{\bf p}_{\pi}),
$$
\begin{equation}
\frac{d\sigma_{{\pi^-}n\to {\pi^-}X}(\sqrt{s},<m_N^{**}>,{\bf p}_{\pi})}
{d{\bf p}_{\pi}}=
\frac{d\sigma_{{\pi^-}n\to {\pi^-}n}(\sqrt{s},<m_N^{**}>,{\bf p}_{\pi})}
{d{\bf p}_{\pi}}
\end{equation}
$$
+
\left[\sigma_{{\pi^-}n \to {\pi^-}{\pi^0}n}(\sqrt{s})+
2\sigma_{{\pi^-}n \to {\pi^-}{\pi^-}p}(\sqrt{s})\right]f_3(\sqrt{s},<m_N^{**}>,{\bf p}_{\pi});
$$
\begin{equation}
\frac{d\sigma_{{\pi^-}p\to {\pi^0}X}(\sqrt{s},<m_N^{**}>,{\bf p}_{\pi})}
{d{\bf p}_{\pi}}=
\sigma_{{\pi^-}p \to {\pi^-}{\pi^0}p}(\sqrt{s})f_3(\sqrt{s},<m_N^{**}>,{\bf p}_{\pi}),
\end{equation}
\begin{equation}
\frac{d\sigma_{{\pi^-}n\to {\pi^0}X}(\sqrt{s},<m_N^{**}>,{\bf p}_{\pi})}
{d{\bf p}_{\pi}}=
\sigma_{{\pi^-}n \to {\pi^-}{\pi^0}n}(\sqrt{s})f_3(\sqrt{s},<m_N^{**}>,{\bf p}_{\pi});
\end{equation}
\begin{equation}
\frac{d\sigma_{{\pi^-}p\to {\pi^+}X}(\sqrt{s},<m_N^{**}>,{\bf p}_{\pi})}
{d{\bf p}_{\pi}}=
\sigma_{{\pi^-}p \to {\pi^+}{\pi^-}n}(\sqrt{s})f_3(\sqrt{s},<m_N^{**}>,{\bf p}_{\pi}),
\end{equation}
\begin{equation}
\frac{d\sigma_{{\pi^-}n\to {\pi^+}X}(\sqrt{s},<m_N^{**}>,{\bf p}_{\pi})}
{d{\bf p}_{\pi}}=0,
\end{equation}
where
\begin{equation}
\frac{d\sigma_{{\pi^-}N\to {\pi^-}N}(\sqrt{s},<m_N^{**}>,{\bf p}_{\pi})}
{d{\bf p}_{\pi}}=
\frac{\pi}{I_2[s,<m^{**}_N>,m_{\pi}]E_{\pi}}
\end{equation}
$$
\times
\frac{\sigma_{{\pi^{-}}N \to {\pi^-}N}(\sqrt{s})}
{4{\pi}}\frac{1}{(\omega_0+E_t)}
$$
$$
\times
\delta\left[\omega_0+E_t-\sqrt{(<m^{**}_{N}>)^2+({\bf Q}_0+{\bf p}_t)^2}\right]
\theta[|{\bf Q}_0+{\bf p}_t|-{\bar p}_F],
$$
\begin{equation}
\omega_0=E_{\pi^-}-E_{\pi}, \,\,\,\,{\bf Q}_0={\bf p}_{\pi^-}-{\bf p}_{\pi}
\end{equation}
and
\begin{equation}
f_3(\sqrt{s},<m_N^{**}>,{\bf p}_{\pi})=
\frac{\pi}{4I_3[s,<m^{**}_N>,m_{\pi},m_{\pi}]E_{\pi}}
\frac{\lambda[s_{N\pi},(<m_N^{**}>)^2,m_{\pi}^2]}{s_{N\pi}}F_{\rm block},
\end{equation}
\begin{equation}
I_{3}[s,<m^{**}_{N}>,m_{\pi},m_{\pi}]=(\frac{{\pi}}{2})^2
\int\limits_{(<m^{**}_{N}>+m_{\pi})^2}^{({\sqrt{s}}-m_{\pi})^2}
\frac{\lambda[x,(<m^{**}_{N}>)^{2},m_{\pi}^2]}{x}
\end{equation}
$$
\times
\frac{\lambda[s,x,m_{\pi}^{2}]}{s}\,dx,
$$
\begin{equation}
s_{N\pi}=s+m_{\pi}^2-2(E_{\pi^-}+E_t)E_{\pi}+2({\bf p}_{\pi^-}+{\bf p}_t){\bf p}_{\pi}.
\end{equation}
Here, $\sigma_{{\pi^-}N\to {\pi^-}N}(\sqrt{s})$ ($N=p,n$) and
$\sigma_{{\pi^-}p\to {\pi^-}{\pi^0}p}(\sqrt{s})$,
$\sigma_{{\pi^-}p\to {\pi^+}{\pi^-}n}(\sqrt{s})$,
$\sigma_{{\pi^-}n\to {\pi^-}{\pi^0}n}(\sqrt{s})$,\\
$\sigma_{{\pi^-}n\to {\pi^-}{\pi^-}p}(\sqrt{s})$
are, respectively, the free total cross sections of the reactions (23), (24) and (25)--(28) calculated
at the ``off-shell" center-of-mass energy $\sqrt{s}$, accessible in these reactions proceeding inside
the nucleus; $\theta(x)$ is the
standard step function, and $F_{\rm block}$ is the Pauli blocking factor, which is given by [53]:
\begin{equation}
F_{\rm block}=\left\{
\begin{array}{lll}
	1
	&\mbox{for ${\bar E}_F \le E_{N}^-$}, \\
	&\\
        \frac{(E_N^+-{\bar E}_F)}{(E_N^+-E_N^-)}
	&\mbox{for $E_N^- < {\bar E}_{F} \le E_N^+$}, \\
    &\\
    0
    &\mbox{for ${\bar E}_{F} > E_N^+$},
\end{array}
\right.	
\end{equation}
where
\begin{equation}
E_N^{\pm}=\frac{{\tilde\omega}\beta\pm{\tilde Q}
\lambda[s_{N\pi},(<m^{**}_{N}>)^{2},m_{\pi}^2]}{2s_{N\pi}}
\end{equation}
and
\begin{equation}
{\tilde\omega}=E_{\pi^-}+E_t-E_{\pi}, \,\,\,\,{\tilde Q}=|{\bf p}_{\pi^-}+{\bf p}_t-{\bf p}_{\pi}|,
\end{equation}
\begin{equation}
\beta=s_{N\pi}+(<m^{**}_{N}>)^{2}-m_{\pi}^2, \,\,\,\,{\bar E}_F=\sqrt{(<m^{**}_{N}>)^{2}+{\bar p}_F^2}.
\end{equation}
Here, the average Fermi momentum ${\bar p}_F$ of the specific target nucleus is defined as [54]:
\begin{equation}
{\bar p}_F=\left(\frac{3{\pi^2}<\rho_N>}{2}\right)^{1/3}.
\end{equation}
For $<\rho_N>=0.55\rho_0$ ($^{12}$C target nucleus), $<\rho_N>=0.76\rho_0$ ($^{184}$W target nucleus) and
$\rho_0=0.16$ fm$^{-3}$, Eq. (52) leads to ${\bar p}_F=216$ MeV/c for $^{12}$C and
${\bar p}_F=241$ MeV/c for $^{184}$W. We will use these magnitudes throughout our calculations.
For the free total cross sections
$\sigma_{{\pi^-}N\to {\pi^-}N}$ ($N=p,n$) and
$\sigma_{{\pi^-}p\to {\pi^-}{\pi^0}p}$,
$\sigma_{{\pi^-}p\to {\pi^+}{\pi^-}n}$,
$\sigma_{{\pi^-}n\to {\pi^-}{\pi^0}n}$,
$\sigma_{{\pi^-}n\to {\pi^-}{\pi^-}p}$
we have employed the following parametrizations:
\begin{figure}[htb]
\begin{center}
\includegraphics[width=12.0cm]{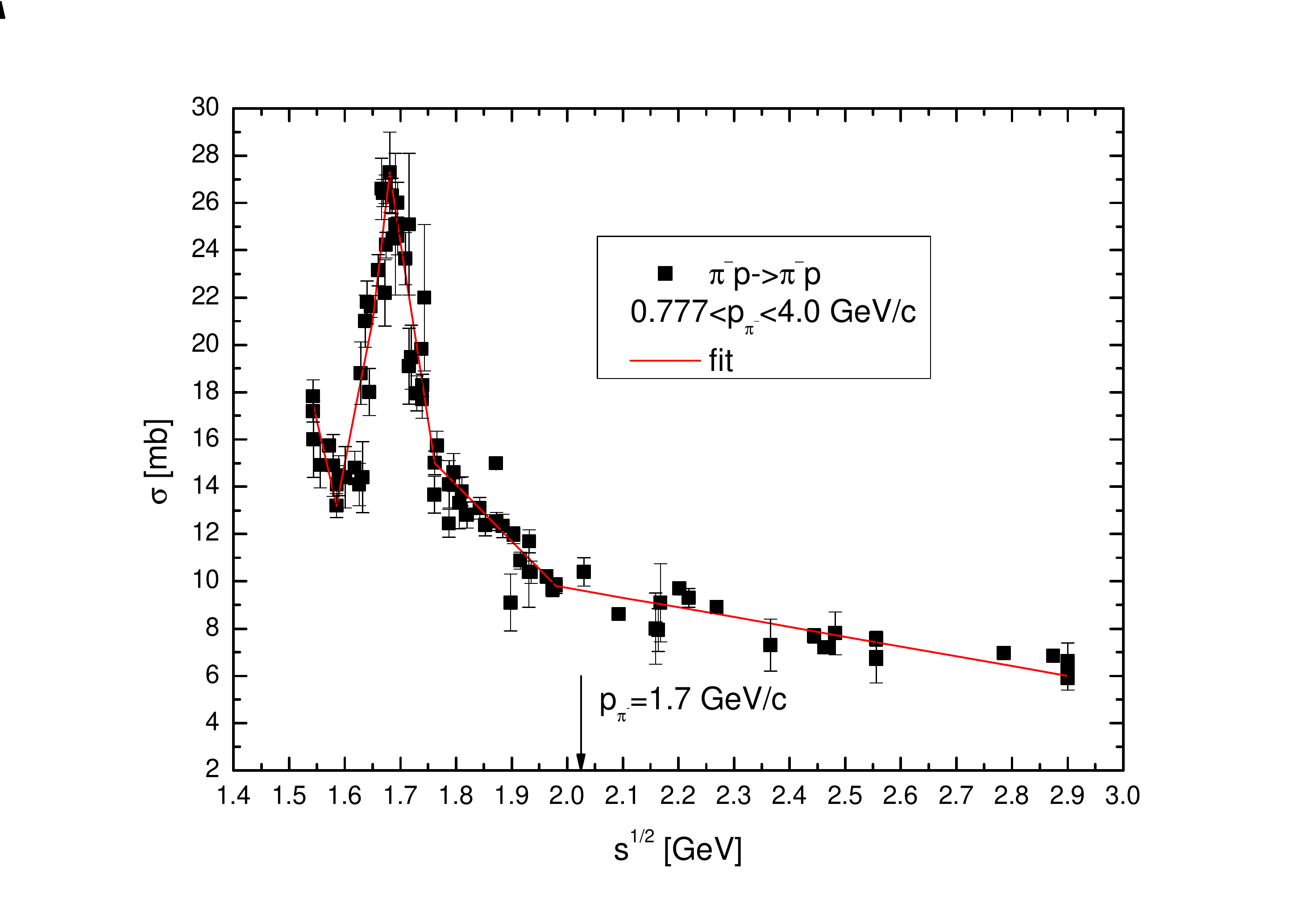}
\vspace*{-2mm} \caption{(color online) The total cross section for the reaction $\pi^-p \to \pi^-p$
as a function of the center-of-mass energy $s^{1/2}$. For notation see the text.}
\label{void}
\end{center}
\end{figure}
\begin{figure}[htb]
\begin{center}
\includegraphics[width=12.0cm]{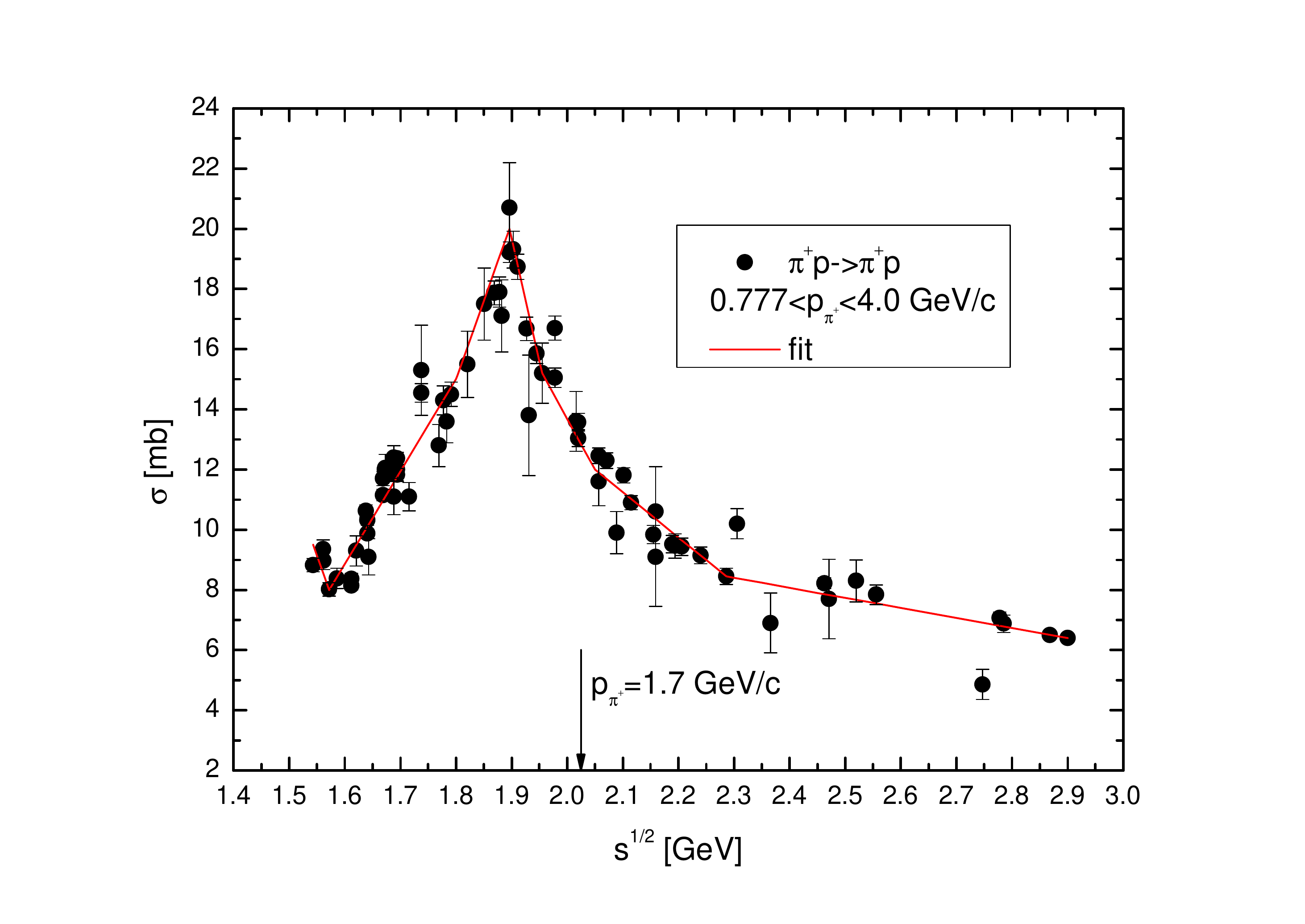}
\vspace*{-2mm} \caption{(color online) The total cross section for the reaction $\pi^+p \to \pi^+p$
as a function of the center-of-mass energy $s^{1/2}$. For notation see the text. Due to isospin symmetry,
this cross section is equal to that of the channel $\pi^-n \to \pi^-n$.}
\label{void}
\end{center}
\end{figure}
\begin{figure}[htb]
\begin{center}
\includegraphics[width=12.0cm]{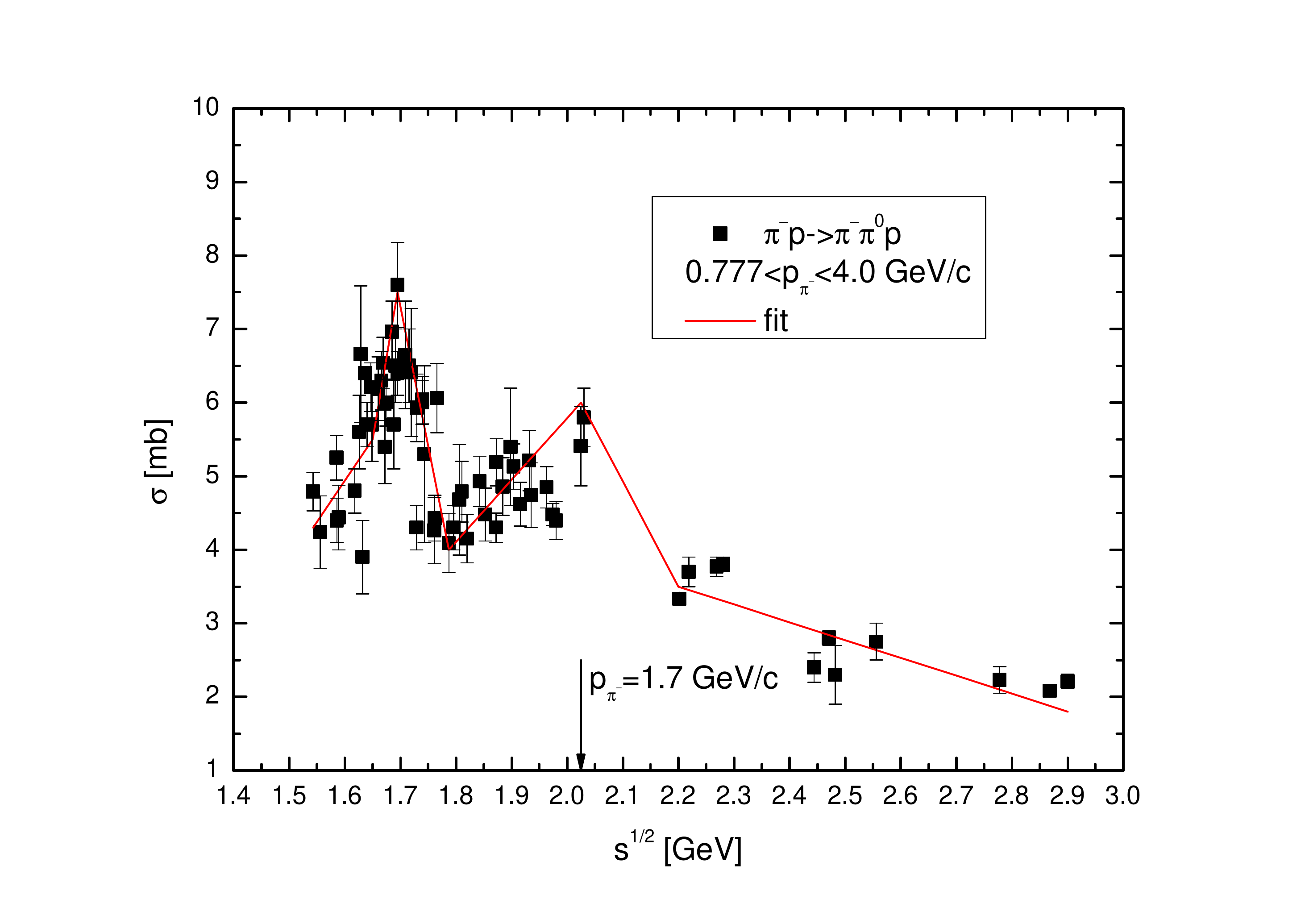}
\vspace*{-2mm} \caption{(color online) The total cross section for the reaction $\pi^-p \to \pi^-\pi^0p$
as a function of the center-of-mass energy $s^{1/2}$. For notation see the text.}
\label{void}
\end{center}
\end{figure}
\begin{figure}[!h]
\begin{center}
\includegraphics[width=12.0cm]{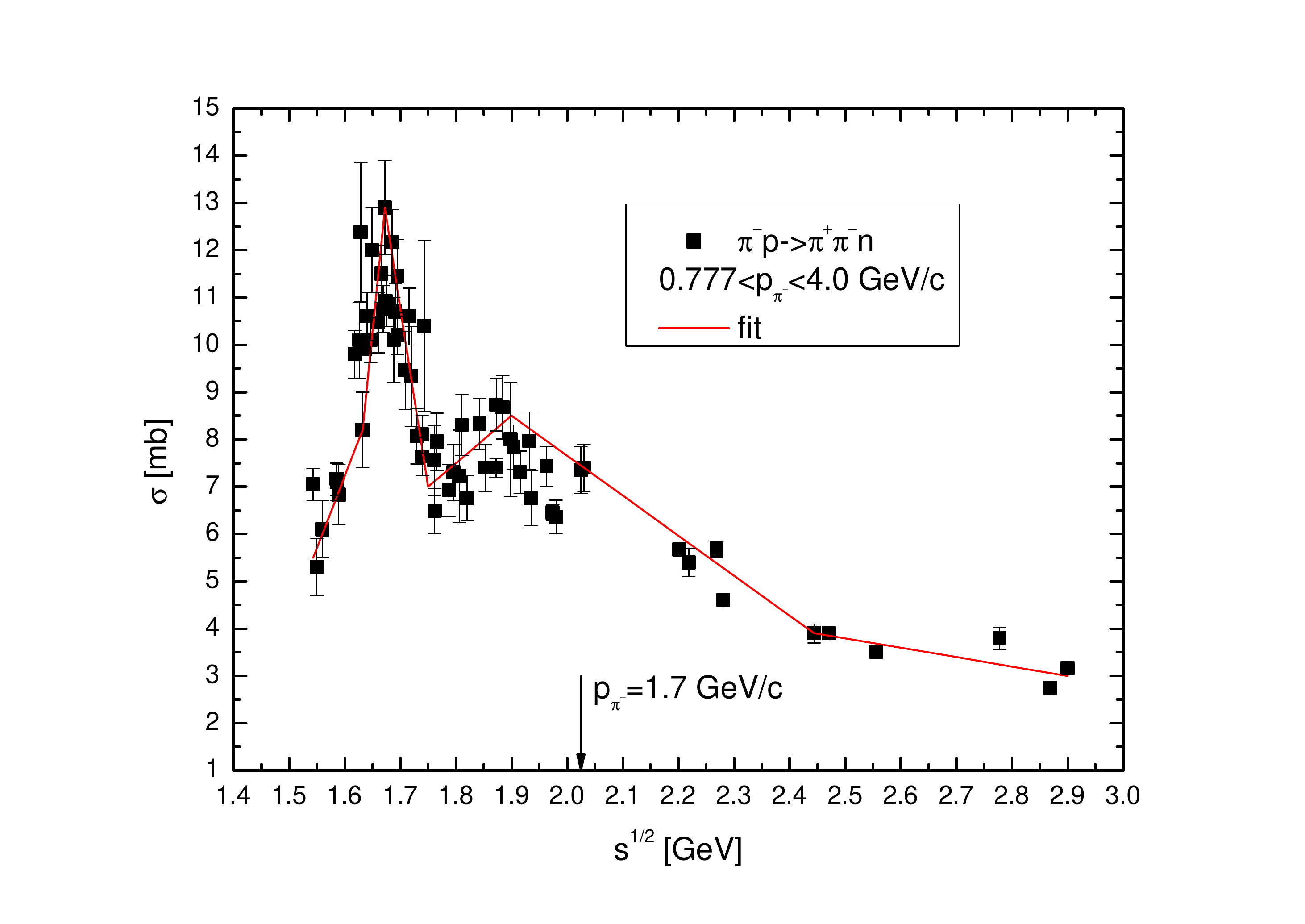}
\vspace*{-2mm} \caption{(color online) The total cross section for the reaction $\pi^-p \to \pi^+\pi^-n$
as a function of the center-of-mass energy $s^{1/2}$. For notation see the text.}
\label{void}
\end{center}
\end{figure}
\begin{figure}[!h]
\begin{center}
\includegraphics[width=12.0cm]{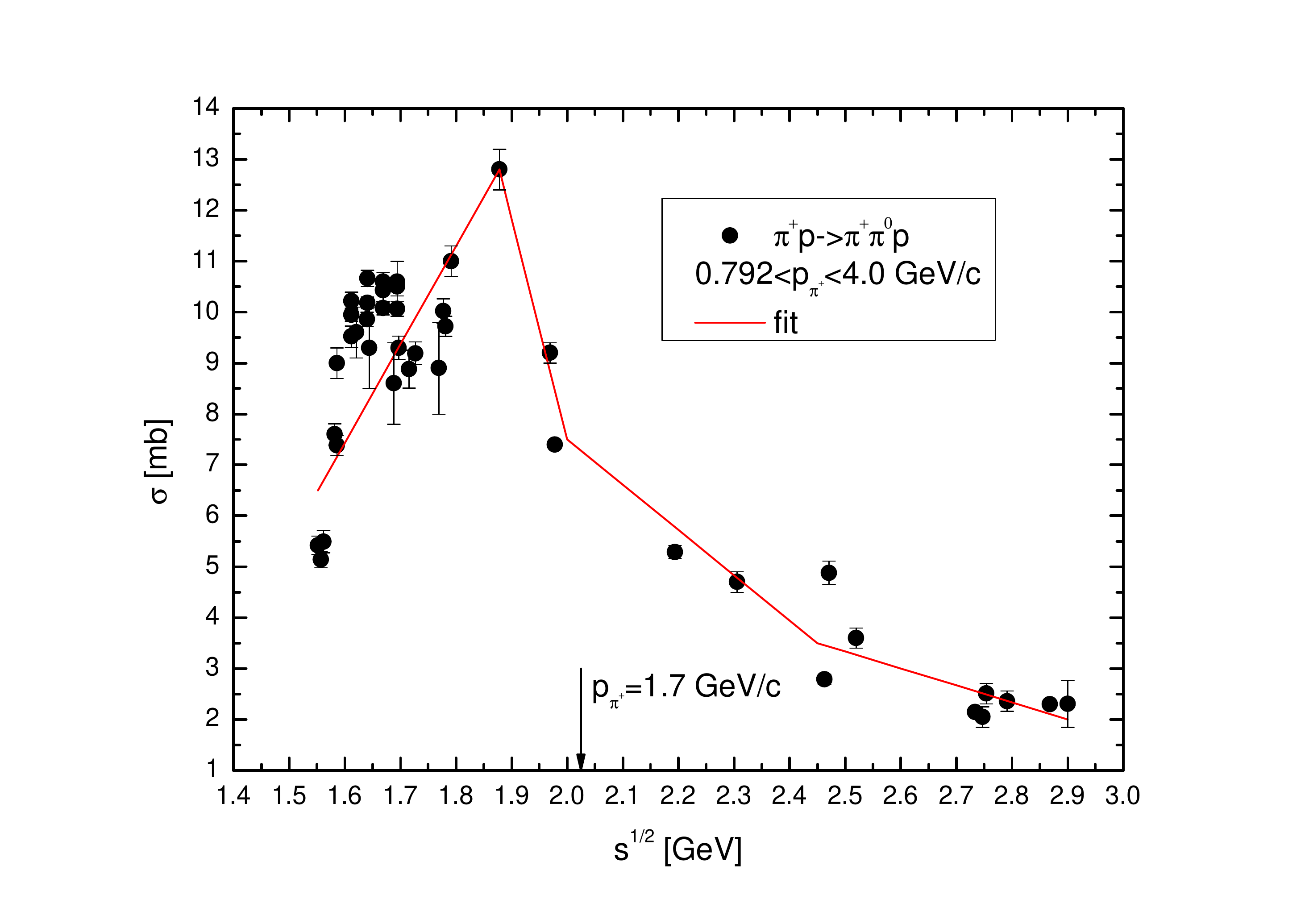}
\vspace*{-2mm} \caption{(color online) The total cross section for the reaction $\pi^+p \to \pi^+\pi^0p$
as a function of the center-of-mass energy $s^{1/2}$. For notation see the text. Due to isospin symmetry,
this cross section is equal to that of the channel $\pi^-n \to \pi^-\pi^0n$.}
\label{void}
\end{center}
\end{figure}
\begin{figure}[!h]
\begin{center}
\includegraphics[width=12.0cm]{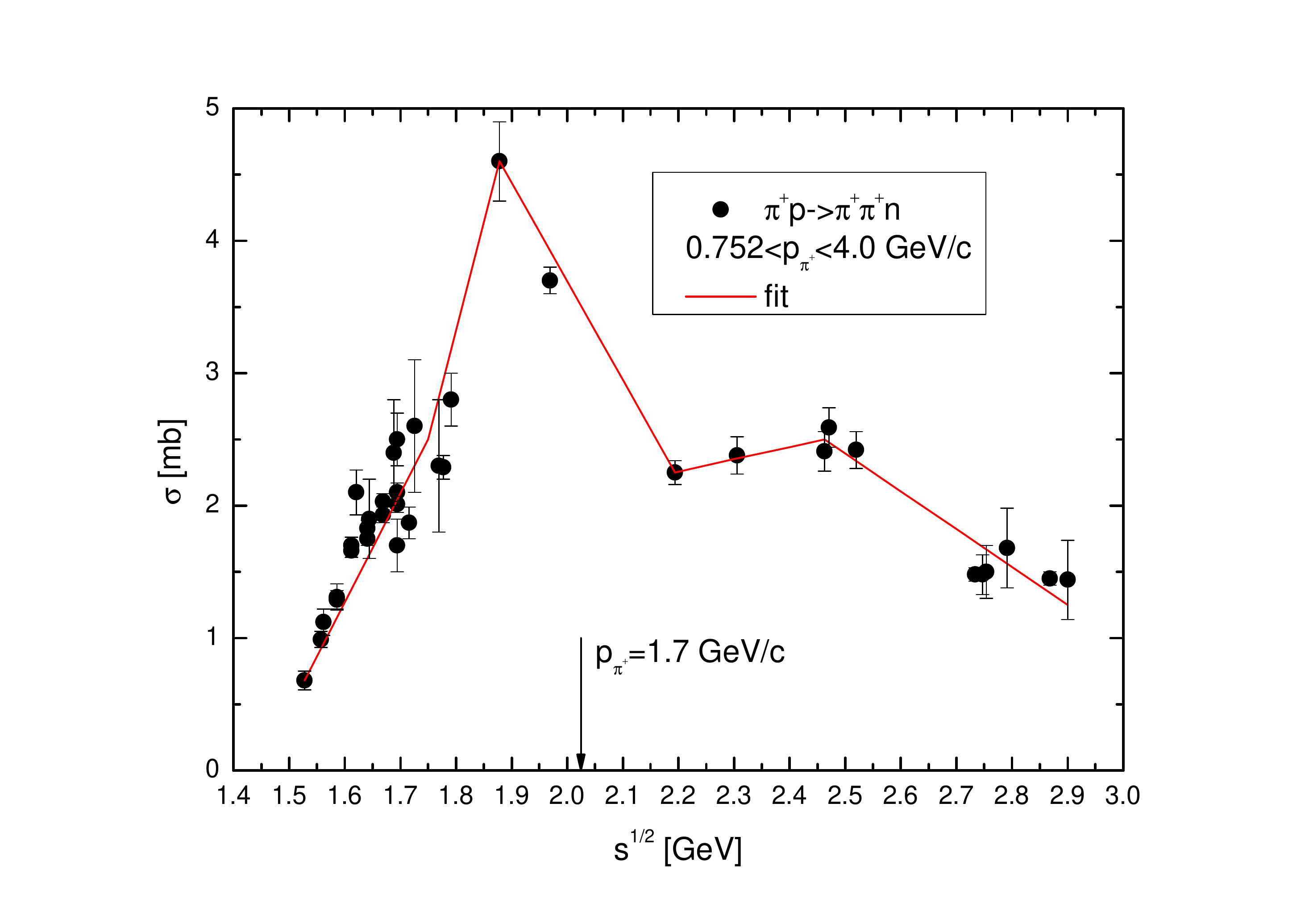}
\vspace*{-2mm} \caption{(color online) The total cross section for the reaction $\pi^+p \to \pi^+\pi^+n$
as a function of the center-of-mass energy $s^{1/2}$. For notation see the text. Due to isospin symmetry,
this cross section is equal to that of the channel $\pi^-n \to \pi^-\pi^-p$.}
\label{void}
\end{center}
\end{figure}
\begin{equation}
\sigma_{{\pi^-}p\to {\pi^-}p}(\sqrt{s})=\left\{
\begin{array}{llllll}
	175.474-102.381(\sqrt{s}/GeV)~[{\rm mb}]
	&\mbox{for $1.543~{\rm GeV} \le \sqrt{s} \le 1.585~{\rm GeV}$}, \\
	&\\
        -177+120(\sqrt{s}/{\rm GeV})~[{\rm mb}]
	&\mbox{for $1.585~{\rm GeV} < \sqrt{s} \le 1.65~{\rm GeV}$}, \\
    &\\
    -314.323+203.226(\sqrt{s}/{\rm GeV})~[{\rm mb}]
    &\mbox{for $1.65~{\rm GeV} < \sqrt{s} \le 1.681~{\rm GeV}$},\\
    &\\
     282.563-151.852(\sqrt{s}/{\rm GeV})~[{\rm mb}]
    &\mbox{for $1.681~{\rm GeV} < \sqrt{s} \le 1.762~{\rm GeV}$},\\
    &\\
     57.029-23.853(\sqrt{s}/{\rm GeV})~[{\rm mb}]
    &\mbox{for $1.762~{\rm GeV} < \sqrt{s} \le 1.98~{\rm GeV}$},\\
    &\\
     17.977-4.13(\sqrt{s}/{\rm GeV})~[{\rm mb}]
    &\mbox{for $1.98~{\rm GeV} < \sqrt{s} \le 2.9~{\rm GeV}$};
\end{array}
\right.	
\end{equation}
\begin{equation}
\sigma_{{\pi^+}p\to {\pi^+}p}(\sqrt{s})
=\sigma_{{\pi^-}n\to {\pi^-}n}(\sqrt{s})
\end{equation}
$$
=\left\{
\begin{array}{llllll}
	89.310-51.724(\sqrt{s}/GeV)~[{\rm mb}]
	&\mbox{for $1.543~{\rm GeV} \le \sqrt{s} \le 1.572~{\rm GeV}$}, \\
	&\\
        -40.264+30.702(\sqrt{s}/{\rm GeV})~[{\rm mb}]
	&\mbox{for $1.572~{\rm GeV} < \sqrt{s} \le 1.8~{\rm GeV}$}, \\
    &\\
    -78.750+52.083(\sqrt{s}/{\rm GeV})~[{\rm mb}]
    &\mbox{for $1.8~{\rm GeV} < \sqrt{s} \le 1.896~{\rm GeV}$},\\
    &\\
     174.251-81.356(\sqrt{s}/{\rm GeV})~[{\rm mb}]
    &\mbox{for $1.896~{\rm GeV} < \sqrt{s} \le 1.955~{\rm GeV}$},\\
    &\\
     81.052-33.684(\sqrt{s}/{\rm GeV})~[{\rm mb}]
    &\mbox{for $1.955~{\rm GeV} < \sqrt{s} \le 2.05~{\rm GeV}$},\\
    &\\
     42.838-15.043(\sqrt{s}/{\rm GeV})~[{\rm mb}]
    &\mbox{for $2.05~{\rm GeV} < \sqrt{s} \le 2.286~{\rm GeV}$},\\
    &\\
     16.083-3.339(\sqrt{s}/{\rm GeV})~[{\rm mb}]
    &\mbox{for $2.286~{\rm GeV} < \sqrt{s} \le 2.9~{\rm GeV}$};
\end{array}
\right.	
$$
\begin{equation}
\sigma_{{\pi^-}p\to {\pi^-}{\pi^0}p}(\sqrt{s})=\left\{
\begin{array}{llllll}
	-13.005+11.215(\sqrt{s}/GeV)~[{\rm mb}]
	&\mbox{for $1.543~{\rm GeV} \le \sqrt{s} \le 1.65~{\rm GeV}$}, \\
	&\\
        -67.833+44.444(\sqrt{s}/{\rm GeV})~[{\rm mb}]
	&\mbox{for $1.65~{\rm GeV} < \sqrt{s} \le 1.695~{\rm GeV}$}, \\
    &\\
     71.985-38.044(\sqrt{s}/{\rm GeV})~[{\rm mb}]
    &\mbox{for $1.695~{\rm GeV} < \sqrt{s} \le 1.787~{\rm GeV}$},\\
    &\\
     -11.016+8.403(\sqrt{s}/{\rm GeV})~[{\rm mb}]
    &\mbox{for $1.787~{\rm GeV} < \sqrt{s} \le 2.025~{\rm GeV}$},\\
    &\\
     34.929-14.286(\sqrt{s}/{\rm GeV})~[{\rm mb}]
    &\mbox{for $2.025~{\rm GeV} < \sqrt{s} \le 2.2~{\rm GeV}$},\\
    &\\
     8.844-2.429(\sqrt{s}/{\rm GeV})~[{\rm mb}]
    &\mbox{for $2.2~{\rm GeV} < \sqrt{s} \le 2.9~{\rm GeV}$};
\end{array}
\right.	
\end{equation}
\begin{equation}
\sigma_{{\pi^-}p\to {\pi^+}{\pi^-}n}(\sqrt{s})=\left\{
\begin{array}{llllll}
	-41.31+30.337(\sqrt{s}/GeV)~[{\rm mb}]
	&\mbox{for $1.543~{\rm GeV} \le \sqrt{s} \le 1.632~{\rm GeV}$}, \\
	&\\
        -183.56+117.5(\sqrt{s}/{\rm GeV})~[{\rm mb}]
	&\mbox{for $1.632~{\rm GeV} < \sqrt{s} \le 1.672~{\rm GeV}$}, \\
    &\\
     139.372-75.641(\sqrt{s}/{\rm GeV})~[{\rm mb}]
    &\mbox{for $1.672~{\rm GeV} < \sqrt{s} \le 1.75~{\rm GeV}$},\\
    &\\
     -10.5+10(\sqrt{s}/{\rm GeV})~[{\rm mb}]
    &\mbox{for $1.75~{\rm GeV} < \sqrt{s} \le 1.9~{\rm GeV}$},\\
    &\\
     24.567-8.456(\sqrt{s}/{\rm GeV})~[{\rm mb}]
    &\mbox{for $1.9~{\rm GeV} < \sqrt{s} \le 2.444~{\rm GeV}$},\\
    &\\
     8.725-1.974(\sqrt{s}/{\rm GeV})~[{\rm mb}]
    &\mbox{for $2.444~{\rm GeV} < \sqrt{s} \le 2.9~{\rm GeV}$};
\end{array}
\right.	
\end{equation}
\begin{equation}
\sigma_{{\pi^+}p\to {\pi^+}{\pi^0}p}(\sqrt{s})=\sigma_{{\pi^-}n\to {\pi^-}{\pi^0}n}(\sqrt{s})
\end{equation}
$$
=\left\{
\begin{array}{llllll}
	-23.492+19.325(\sqrt{s}/GeV)~[{\rm mb}]
	&\mbox{for $1.552~{\rm GeV} \le \sqrt{s} \le 1.878~{\rm GeV}$}, \\
	&\\
        94.386-43.443(\sqrt{s}/{\rm GeV})~[{\rm mb}]
	&\mbox{for $1.878~{\rm GeV} < \sqrt{s} \le 2.0~{\rm GeV}$}, \\
    &\\
     25.278-8.889(\sqrt{s}/{\rm GeV})~[{\rm mb}]
    &\mbox{for $2.0~{\rm GeV} < \sqrt{s} \le 2.45~{\rm GeV}$},\\
    &\\
     11.666-3.333(\sqrt{s}/{\rm GeV})~[{\rm mb}]
    &\mbox{for $2.45~{\rm GeV} < \sqrt{s} \le 2.9~{\rm GeV}$};
\end{array}
\right.	
$$
\begin{equation}
\sigma_{{\pi^+}p\to {\pi^+}{\pi^+}n}(\sqrt{s})=\sigma_{{\pi^-}n\to {\pi^-}{\pi^-}p}(\sqrt{s})
\end{equation}
$$
=\left\{
\begin{array}{llllll}
	-11.847+8.198(\sqrt{s}/GeV)~[{\rm mb}]
	&\mbox{for $1.528~{\rm GeV} \le \sqrt{s} \le 1.75~{\rm GeV}$}, \\
	&\\
        -26.211+16.406(\sqrt{s}/{\rm GeV})~[{\rm mb}]
	&\mbox{for $1.75~{\rm GeV} < \sqrt{s} \le 1.878~{\rm GeV}$}, \\
    &\\
     18.567-7.437(\sqrt{s}/{\rm GeV})~[{\rm mb}]
    &\mbox{for $1.878~{\rm GeV} < \sqrt{s} \le 2.194~{\rm GeV}$},\\
    &\\
     0.212+0.929(\sqrt{s}/{\rm GeV})~[{\rm mb}]
    &\mbox{for $2.194~{\rm GeV} < \sqrt{s} \le 2.463~{\rm GeV}$},\\
    &\\
     9.544-2.86(\sqrt{s}/{\rm GeV})~[{\rm mb}]
    &\mbox{for $2.463~{\rm GeV} < \sqrt{s} \le 2.9~{\rm GeV}$}.
\end{array}
\right.	
$$
We show in Figs. 1--6
a comparison of the results of calculations using Eqs.~(53)--(58) (solid lines) with the available
experimental data [49]. The arrows here indicate the collision energy $\sqrt{s}$
corresponding to the incident pion momentum of 1.7 GeV/c and a free target nucleon at rest. It can be seen that
the parametrizations (53)--(58) fit  the existing sets of data at center-of-mass energies $\sqrt{s}$ well,
accessible in the calculation of ${\pi}N$ and $2{\pi}N$ production in ${\pi^-}N$ interactions at the beam momentum
of interest with allowance for the Fermi motion and nuclear binding of intranuclear nucleons.

   The differential cross section for $\phi$ meson production in ${\pi^-}A$ collisions
from the two-step processes (23), (24), (1) and (25)--(28), (1), (29)--(31),
corresponding to the kinematical conditions of the HADES experiment,
can be determined analogously to Eq. (21) as:
\begin{equation}
\frac{d\sigma_{{\pi^-}A\to {\phi}X}^{({\rm sec})}
(p_{\pi^-},p_{\phi})}{dp_{\phi}}=\int\limits_{{\Delta}{\bf \Omega}_{\phi}}^{}d{\bf \Omega}_{\phi}
\frac{d\sigma_{{\pi^-}A\to {\phi}X}^{({\rm sec})}
({\bf p}_{\pi^-},{\bf p}_{\phi})}{d{\bf p}_{\phi}}p_{\phi}^2.
\end{equation}

Now, we discuss the results of calculations for $\phi$ production in ${\pi^-}A$ reactions
within the model outlined above.
\begin{figure}[!h]
\begin{center}
\includegraphics[width=16.0cm]{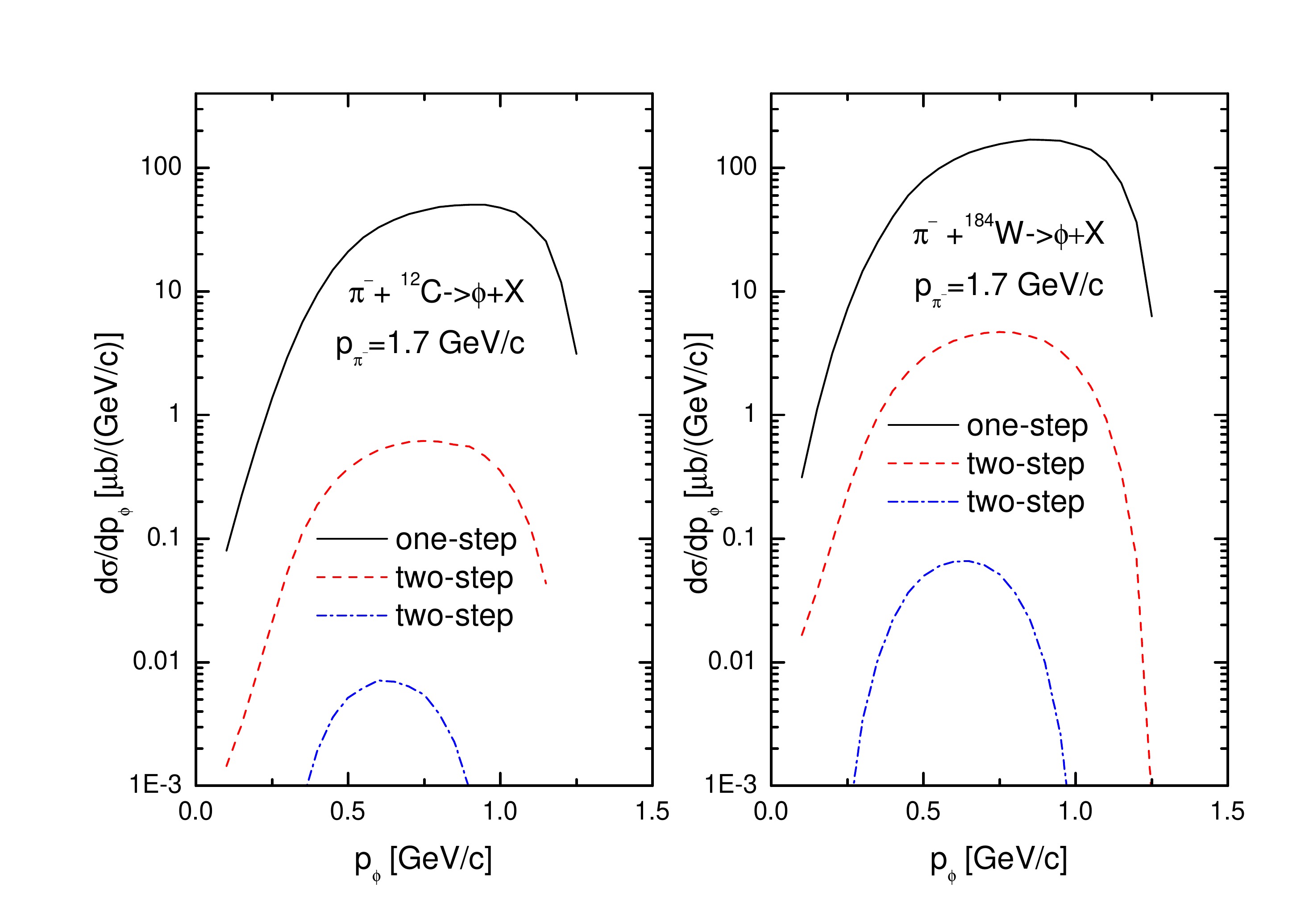}
\vspace*{-2mm} \caption{(color online) Momentum differential cross sections for the production of $\phi$
mesons from one-step (1) (solid lines), two-step (23), (24), (1) (dashed lines) and two-step (25)--(28),
(1), (29)--(31) (dotted-dashed lines) processes in the laboratory polar angular range of
10$^{\circ}$--45$^{\circ}$ in the interaction of $\pi^-$ mesons of momentum of 1.7 GeV/c with $^{12}$C
(left panel) and $^{184}$W (right panel) nuclei for ${\phi}N$ absorption cross section of 10 mb and
effective scalar potentials at density $\rho_0$ of -20 and 25, -60 MeV, seen by the
$\phi$ mesons and secondary nucleons,
created in the channels (1), (29)--(31) and (23), (24), (25)--(28), respectively.}
\label{void}
\end{center}
\end{figure}
\begin{figure}[!h]
\begin{center}
\includegraphics[width=16.0cm]{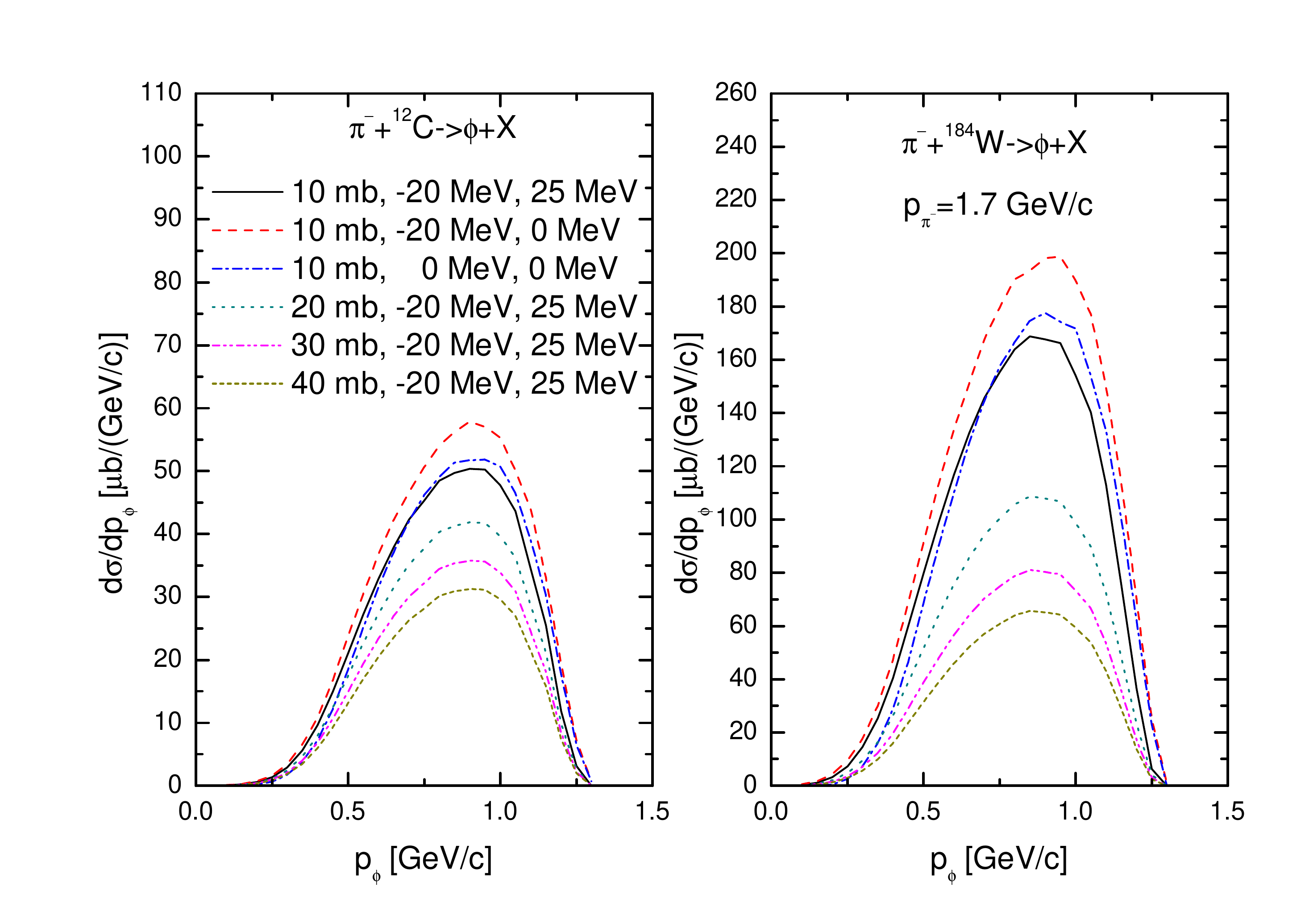}
\vspace*{-2mm} \caption{(color online) Momentum differential cross sections for the production of $\phi$
mesons from the primary ${\pi^-}p \to {\phi}n$ channel in the laboratory polar angular range of
10$^{\circ}$--45$^{\circ}$ in the interaction of $\pi^-$ mesons of momentum of 1.7 GeV/c with $^{12}$C
(left panel) and $^{184}$W (right panel) nuclei for different values of the ${\phi}N$ absorption cross
section as well as $\phi$ meson and secondary neutron effective scalar potentials at density $\rho_0$
indicated in the inset.}
\label{void}
\end{center}
\end{figure}
\begin{figure}[!h]
\begin{center}
\includegraphics[width=16.0cm]{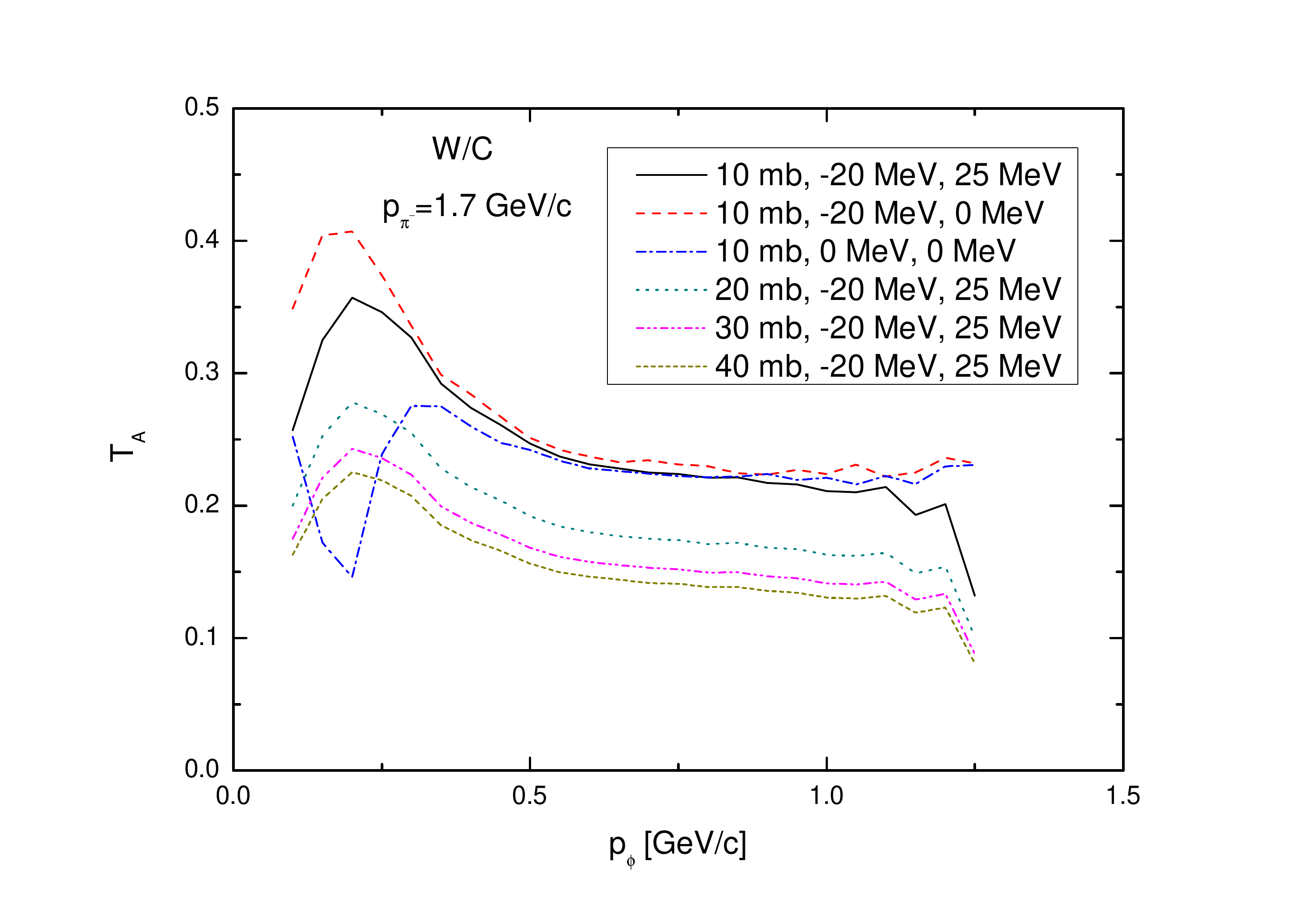}
\vspace*{-2mm} \caption{(color online) Transparency ratio $T_A$ as a function of the $\phi$ momentum
for combination $^{184}$W/$^{12}$C as well as for the $\phi$ laboratory polar angular range of
10$^{\circ}$--45$^{\circ}$, for an incident $\pi^-$ meson momentum of 1.7 GeV/c and for
different values of the ${\phi}N$ absorption cross
section and $\phi$ meson and secondary neutron effective scalar potentials at density $\rho_0$
indicated in the inset.}
\label{void}
\end{center}
\end{figure}

\section*{3 Results and discussion}

\hspace{0.5cm} First, we consider the momentum dependence of the absolute
$\phi$ production cross sections from the one-step (1) as well as from the
two-step (23), (24), (1) and (25)--(28), (1), (29)--(31)
$\phi$ creation mechanisms in ${\pi^-}$$^{12}$C,  ${\pi^-}$$^{184}$W collisions, calculated on the
basis of Eqs. (21), (32)--(59) for incident pion momentum of 1.7 GeV/c, for laboratory $\phi$ production
angles of 10$^{\circ}$--45$^{\circ}$ and for ${\phi}N$ absorption cross section of 10 mb.
The in-medium mass shifts at saturation density $\rho_0$ for the $\phi$ meson as well as for the secondary
nucleons, produced in the channels (1), (29)--(31) and (23), (24), (25)--(28), were chosen in the
calculations as -20, 25 and -60 MeV, respectively. These dependences are shown in Fig.~7.
The primary ${\pi^-}p \to {\phi}n$ channel plays the dominant role
for the kinematical conditions of the HADES experiment. This gives us confidence that the two-step
processes ${\pi^-}N \to {\pi^-}N$, ${\pi^-}p \to {\phi}n$ and
${\pi^-}N \to {\pi}{\pi}N$, ${\pi}N \to {\phi}N$ can be ignored in our subsequent calculations.
Since in these calculations, contrary to the cases of proton- and photon-induced $\phi$ production in
nuclei, there is at least no need to know the meson creation cross section off the neutron,
strong constraints on the inelastic ${\phi}N$ total cross section $\sigma_{{\phi}N}$
can be made from comparing the results of the present one-step model calculations with the future data from
the HADES experiment.

   Now, we concentrate on the momentum dependence of the absolute $\phi$ meson yield from the primary process
(1) in the HADES acceptance window for the incident pion momentum of 1.7 GeV/c. The momentum differential cross
sections for $\phi$ production from  $^{12}$C and $^{184}$W, calculated on the basis of
Eq.~(21) in the scenarios of collisional broadening of the $\phi$ meson characterized by the values of
$\sigma_{{\phi}N}=10$, 20, 30, 40 mb,
with and without accounting for in-medium $\phi$ and secondary neutron mass shifts of -20 and 25 MeV, respectively,
are depicted in Fig.~8. These values are motivated by the results from the
$\phi$ production experiments [17--20].
It is seen that these cross sections reveal some sensitivity to the
above shifts in the momentum range of 0.6--1.1 GeV/c (where they are the greatest)
studied in the HADES experiment. Thus, the inclusion of the in-medium $\phi$ mass shift of -20 MeV at normal
nuclear matter density alone leads to an enhancement of the $\phi$ production cross sections here by about a
factor of 1.1--1.3 as compared to those obtained in the scenario without taking into account both this shift
and that of the secondary neutron, created together with the $\phi$ meson in the primary channel (1). Additional inclusion
of the latter shift of 25 MeV at saturation density $\rho_0$ results in a reduction of the $\phi$
yields from $^{12}$C and $^{184}$W target nuclei by about the same factor, which indicates
that only simultaneous application of the $\phi$ meson and final neutron effective scalar potentials leaves
these yields practically unaffected (compare the solid and dotted-dashed lines in Fig.~8). Looking at this figure, we
see that the obtained results also depend strongly  on the $\phi$--nucleon absorption cross section,
especially for the heavy target nucleus $^{184}$W. We observe differences
($\sim$ 15--20\% for $^{12}$C and $\sim$ 20--40\% for $^{184}$W) between all calculations
corresponding to different options for the ${\phi}N$ absorption cross section. However, these differences are comparable
to those caused by the modifications of the $\phi$ meson and secondary neutron masses in nuclear matter.
Therefore, these modifications can mask the determination of the ${\phi}N$ absorption cross section from the
absolute $\phi$ momentum distribution measurements at initial pion momentum of 1.7 GeV/c. In this context
it is interesting to consider the possibility of extracting the above cross section from the transparency ratio
measurements, where the role of meson in-medium modification is expected to be negligible [39, 55].

   Figure~9 shows the momentum dependence of the transparency ratio $T_A$ for W/C combination for $\phi$ mesons
produced in the primary ${\pi^-}p \to {\phi}n$ reaction channel at 10$^{\circ}$--45$^{\circ}$ laboratory angles
by 1.7 GeV/c $\pi^-$ mesons. The transparency ratio is calculated on the basis of Eq.~(22), accounting for
both the adopted options for the ${\phi}N$ absorption cross section and for the $\phi$ meson and final neutron
in-medium mass shifts. It is seen from this figure that the differences between all calculations corresponding
to different options for these shifts are indeed insignificant at the $\phi$ momenta of 0.6--1.1 GeV/c in which we are interested.
On the other hand, there are sizeable differences between the choices $\sigma_{{\phi}N}=10$ mb and
$\sigma_{{\phi}N}=20$ mb, $\sigma_{{\phi}N}=20$ mb and $\sigma_{{\phi}N}=30$ mb and
$\sigma_{{\phi}N}=30$ mb and $\sigma_{{\phi}N}=40$ mb for $\phi$--nucleon total inelastic cross section
at these momenta. They are $\sim$ 25, 13 and 7\%, respectively. This means that the future precise HADES $\phi$
production data on momentum dependence of the transparency ratio for $\phi$ mesons should help to distinguish
at least between weak ($\sigma_{{\phi}N} \sim 10$ mb), relatively weak ($\sigma_{{\phi}N} \sim 20$ mb) and strong
($\sigma_{{\phi}N} \sim 30$ mb) ${\phi}$ absorption in cold nuclear matter.

   Thus, we come to the conclusion that a relative observable such as the transparency ratio for the $\phi$
mesons considered above has an insignificant sensitivity to the $\phi$ meson and secondary neutron in-medium
mass shifts at the $\phi$ momenta studied in the HADES experiment and, hence, can be useful to help
determine the ${\phi}N$ absorption cross section. Having this cross section fixed, the HADES $\phi$
momentum distribution measurements at incident pion momentum of 1.7 GeV/c will open an opportunity to
shed light on the possible mass shift of the $\phi$ meson in cold nuclear matter.
It should be noted that the J-PARC E16 Collaboration also intends [56]  to study
this shift  more systematically via the $\phi$ dilepton decay channel with statistics two orders of magnitude higher
than the KEK-E325 experiment [16].

\section*{4 Conclusions}

\hspace{0.5cm} With the aim of studying the absorption of $\phi$ mesons in the
nuclear medium, we have developed a spectral function approach
for the description of their production in $\pi^-$ meson-induced nuclear reactions near the threshold.
The approach takes into account both primary $\pi^-$ meson--proton and secondary pion--nucleon $\phi$
production processes as well as four different options for the ${\phi}N$ absorption cross section
$\sigma_{{\phi}N}$ and two different choices for the in-medium mass shifts of the $\phi$ meson
and secondary neutron, produced together with the $\phi$ in the direct reaction channel
${\pi^-}p \to {\phi}n$. We have found that this channel dominates in $\phi$ production off nuclei in
the HADES acceptance window at incident pion momentum of 1.7 GeV/c. We have calculated the momentum
dependence of the absolute and relative (transparency ratio) $\phi$ meson yields from the above direct
channel at this incident pion momentum. It was demonstrated that the transparency ratio for the $\phi$
mesons has, contrary to the absolute cross sections, an insignificant sensitivity to the $\phi$ meson
and secondary neutron in-medium mass shifts at $\phi$ momenta studied in the HADES experiment.
On the other hand, it was shown that there are measurable changes in the transparency ratio
due to the ${\phi}N$ absorption cross section. Hence, this relative observable can be useful
to help determine this cross section from the data taken in the HADES experiment.
\\
\\
{\bf Acknowledgments}
\\
\\
The author is very grateful to Laura Fabbietti and Volker Metag for initiation of this study as well as
to Joana Wirth for performing the ongoing analysis of the pion-induced $\phi$ production data from the
HADES experiment.
\\
\\

\end{document}